\documentstyle{mn}

\newif\ifAMStwofonts

\ifoldfss
  \ifCUPmtlplainloaded \else
    \NewTextAlphabet{textbfit} {cmbxti10} {}
    \NewTextAlphabet{textbfss} {cmssbx10} {}
    \NewMathAlphabet{mathbfit} {cmbxti10} {} 
    \NewMathAlphabet{mathbfss} {cmssbx10} {} 
  \fi
  \ifAMStwofonts
    \ifCUPmtlplainloaded \else
      \NewSymbolFont{upmath} {eurm10}
      \NewSymbolFont{AMSa} {msam10}
      \NewMathSymbol{\upi}     {0}{upmath}{19}
      \NewMathSymbol{\umu}     {0}{upmath}{16}
      \NewMathSymbol{\upartial}{0}{upmath}{40}
      \NewMathSymbol{\leqslant}{3}{AMSa}{36}
      \NewMathSymbol{\geqslant}{3}{AMSa}{3E}

    \fi
  \fi
\fi 

\ifnfssone
  \newmathalphabet{\mathit}
  \addtoversion{normal}{\mathit}{cmr}{m}{it}
  \addtoversion{bold}{\mathit}{cmr}{bx}{it}
  \newmathalphabet{\mathbfit} 
  \addtoversion{normal}{\mathbfit}{cmr}{bx}{it}
  \addtoversion{bold}{\mathbfit}{cmr}{bx}{it}
  \newmathalphabet{\mathbfss} 
  \addtoversion{normal}{\mathbfss}{cmss}{bx}{n}
  \addtoversion{bold}{\mathbfss}{cmss}{bx}{n}
  \ifAMStwofonts
    \ifCUPmtlplainloaded \else
      \UseAMStwoboldmath
      \makeatletter
      \new@mathgroup\upmath@group
      \define@mathgroup\mv@normal\upmath@group{eur}{m}{n}
      \define@mathgroup\mv@bold\upmath@group{eur}{b}{n}
      \edef\UPM{\hexnumber\upmath@group}
      \new@mathgroup\amsa@group
      \define@mathgroup\mv@normal\amsa@group{msa}{m}{n}
      \define@mathgroup\mv@bold\amsa@group{msa}{m}{n}
      \edef\AMSa{\hexnumber\amsa@group}
      \makeatother
      \mathchardef\upi="0\UPM19
      \mathchardef\umu="0\UPM16
      \mathchardef\upartial="0\UPM40
      \mathchardef\leqslant="3\AMSa36
      \mathchardef\geqslant="3\AMSa3E
    \fi
  \fi
\fi 

\ifnfsstwo
  \DeclareMathAlphabet{\mathbfit}{OT1}{cmr}{bx}{it}
  \SetMathAlphabet\mathbfit{bold}{OT1}{cmr}{bx}{it}
  \DeclareMathAlphabet{\mathbfss}{OT1}{cmss}{bx}{n}
  \SetMathAlphabet\mathbfss{bold}{OT1}{cmss}{bx}{n}
  \ifAMStwofonts
    \ifCUPmtlplainloaded \else
      \DeclareSymbolFont{UPM}{U}{eur}{m}{n}
      \SetSymbolFont{UPM}{bold}{U}{eur}{b}{n}
      \DeclareSymbolFont{AMSa}{U}{msa}{m}{n}
      \DeclareMathSymbol{\upi}{0}{UPM}{"19}
      \DeclareMathSymbol{\umu}{0}{UPM}{"16}
      \DeclareMathSymbol{\upartial}{0}{UPM}{"40}
      \DeclareMathSymbol{\leqslant}{3}{AMSa}{"36}
      \DeclareMathSymbol{\geqslant}{3}{AMSa}{"3E}
    \fi
  \fi
\fi 

\ifCUPmtlplainloaded \else
  \ifAMStwofonts \else 
    \def\upi{\pi}
    \def\umu{\mu}
    \def\upartial{\partial}
  \fi
\fi

\title{On the source of the late-time infrared luminosity of SN~1998S
and other type~II supernovae} \author[M. Pozzo, et al.]
{M. Pozzo$^{1}$, W.P.S. Meikle$^{1}$, A. Fassia$^{1}$,
T. Geballe$^{2}$, P. Lundqvist$^{3}$, N.N. Chugai$^{4}$ \newauthor and
J. Sollerman$^{3}$ \\ $^1$ Imperial College London, Blackett
Laboratory, Prince Consort Road, London, SW7 2BW, UK \\ $^2$Gemini
Observatory, 670 N. A'ohoku Place, Hilo, HI 96720, USA\\ $^3$Stockholm
Observatory, AlbaNova, Dept. of Astronomy, Stockholm SE 106 91,
Sweden\\ $^4$Insitute of Astronomy, RAS, Pyatnitskaya 48, 109017,
Russia\\} \date{Accepted ... .  Received .... ; in original form 2004
March 2}

\pagerange{\pageref{firstpage}--\pageref{lastpage}}
\pubyear{0000}

\begin{document}

\maketitle

\label{firstpage}

\begin{abstract}
We present late-time near-infrared (NIR) and optical observations of
the type~IIn SN~1998S. The NIR photometry spans 333--1242~days after
explosion, while the NIR and optical spectra cover 333--1191~days and
305--1093~days respectively. The NIR photometry extends to the
$M'$-band (4.7~$\mu$m), making SN~1998S only the second ever supernova
for which such a long IR wavelength has been detected. The shape and
evolution of the H$\alpha$ and He\,{\sc i} 1.083~$\mu$m line profiles
indicate a powerful interaction with a progenitor wind, as well as
providing evidence of dust condensation within the ejecta.  The latest
optical spectrum suggests that the wind had been flowing for at least
430~years.  The intensity and rise of the $HK$ continuum towards
longer wavelengths together with the relatively bright $L'$ and $M'$
magnitudes shows that the NIR emission was due to hot dust
newly-formed in the ejecta and/or pre-existing dust in the progenitor
circumstellar medium (CSM).  The NIR spectral energy distribution
(SED) at about 1~year is well-described by a single-temperature
blackbody spectrum at about 1200~K. The temperature declines over
subsequent epochs. After $\sim$2~years the blackbody matches are less
successful, probably indicating an increasing range of temperatures in
the emission regions.  Fits to the SEDs achieved with blackbodies
weighted with $\lambda^{-1}$ or $\lambda^{-2}$ emissivity are almost
always less successful. Possible origins for the NIR emission are
considered. Significant radioactive heating of ejecta dust is ruled
out, as is shock/X-ray-precursor heating of CSM dust.  More plausible
sources are (a) an IR-echo from CSM dust driven by the UV/optical peak
luminosity, and (b) emission from newly-condensed dust which
formed within a cool, dense shell produced by the ejecta shock/CSM
interaction. We argue that the evidence favours the condensing dust
hypothesis, although an IR-echo is not ruled out. Within the
condensing-dust scenario, the IR luminosity indicates the presence of
at least 10$^{-3}$~M$_{\odot}$ of dust in the ejecta, and probably
considerably more.  Finally, we show that the late-time $(K-L')_0$
evolution of type~II supernovae may provide a useful tool for
determining the presence or absence of a massive CSM around their
progenitor stars.
\end{abstract}


\section{Introduction}
One of the challenges of supernova research is to obtain evidence
about the nature and environment of the progenitor.  Type~IIn
supernovae (SNe~IIn) are so-called because of the presence of narrow
lines in the spectra, originating in a relatively undisturbed
circumstellar medium (CSM) (Schlegel 1990).  The progenitors must
therefore have undergone one or more mass-loss phases before
explosion.  By using the interaction of the supernova explosion with
the resulting CSM we can obtain clues about the nature of the
progenitor. \\

The study of SNe IIn may also help us to understand the origin of dust
in the universe.  More than 30 years ago, it was suggested (Cernuschi,
Marsicano \& Kimel 1965; Cernuschi, Marsicano \& Codina 1967; Hoyle \&
Wickramasinghe 1970) that supernovae could be important sources of
interstellar dust. More recent studies of the origins of grains (Gehrz
1989; Tielens 1990; Dwek 1998; Todini \& Ferrara 2001; Nozawa et
al. 2003) still invoke core-collapse SNe as significant contributors
of dust to the interstellar medium in the past and present.  Dust
grains associated with SNe can be detected both via their attenuation
effects on optical/near-IR radiation, and by their thermal
re-radiation at longer IR wavelengths of the absorbed
energy. Observation of attenuation provides a relatively unambiguous
way of demonstrating grain condensation in the ejecta. This process is
revealed by the relative suppression of the red wings of the broad
spectral lines.  However, apart from SN~1998S (Leonard et al. 2000;
Gerardy et al. 2000; this work), only for two events, SN~1987A and
SN~1999em, has the condensation of dust in the ejecta been
demonstrated via the attenuation of the red wing of the line profiles
(Danziger et al. 1991; Lucy, Danziger \& Gouiffes 1989; Lucy et
al. 1991; Elmhamdi et al. 2003). \\

The second method of demonstrating grain condensation via the
appearance of strong IR radiation (an `IR excess') as a supernova
ages, is more challenging.  At least 10 SNe have shown a strong,
late-time (i.e. $t>100$~d) NIR excess (relative to that expected by
extrapolation from the optical region; see references in Gerardy et
al. 2002).  The natural interpretation is that this radiation is
produced by hot dust.  However, while such grains may be the result of
condensation processes in the expanding SN ejecta, they may also have
formed in pre-explosion circumstellar material ejected during one or
more mass-loss episodes experienced by the progenitor star. In this
latter case, the IR emission is produced as an ``IR echo'' of the
maximum-light luminosity from the CSM dust.  Thus, in the
interpretation of such strong NIR emission from SNe, a key issue is
the location of the source.  Only in the case of SN~1987A has
conclusive evidence for dust condensation in the ejecta been presented
on the basis of the NIR and mid-IR emission (Moseley et al. 1989;
Whitelock et al. 1989; Suntzeff \& Bouchet 1990; Dwek 1991; Lucy et
al. 1991; Dwek et al. 1992; Meikle et al. 1993; Roche, Aitken \& Smith
1993; Colgan et al. 1994).  Recently, Dunne et al. (2003) have
presented evidence based on sub-mm observations that
$\sim2-4~M_{\odot}$ of cold dust is present in the supernova remnant
(SNR) Cassiopeia~A. They argue that the dust must have been produced
during the explosion. A similar result for the Kepler SNR was
presented by Morgan et al. (2003) although in this case an alternative
CSM origin for the dust was not ruled out, which would be more
consistent with the most commonly believed origin for this supernova,
i.e. a type Ia explosion. Such sub-mm studies may provide important
support for the hypothesis that supernovae are a major source of dust.
However, at the moment they have the disadvantages that the supernova
types are uncertain, and that the number of examples which can be
studied in this way is quite small.  In addition, the results of Dunne
et al. (2003) and Morgan et al. (2003) have been criticised by Dwek
(2004) who points out that, given the mass of condensible elements
expected in the ejecta, a very unlikely condensation efficiency of
100\% would be required to yield even 2~M$_{\odot}$ of dust.  Moreover
thermal sputtering would reduce this mass of dust.  Dwek offers an
alternative scenario where the sub-mm emission is from conductive
needles (''metallic whiskers'') which may be long enough to be very
efficient emitters at sub-mm wavelengths thus requiring a dust mass of
no more than $10^{-3}$~M$_{\odot}$ to account for the observed
radiation.\\

Lastly, we note that the unique wavelength and temporal coverage of
SN~1987A also made it possible to infer dust condensation from (a)
spectroscopic evidence for the temporal depletion of gaseous-phase
elements in the ejecta e.g., [0\,{\sc i}] (6300,~6364~\AA), [Mg\,{\sc
i}] (4571 \AA), Si\,{\sc i} (1.2~$\mu$m), [Si\,{\sc i}] (1.65~$\mu$m)
and [Ca\,{\sc ii}] (7300~\AA) (Lucy et al. 1991; Spyromilio et
al. 1991; Meikle et al. 1993), and (b) analysis of the luminosity
budget of the ejecta (e.g. Whitelock et al. 1989).  As observing
facilities improve, these last two methods of dust detection may
eventually be applicable to SNe at more typical distances.\\

To provide conclusive evidence of SNe as major dust sources, we need
to assemble a statistically significant sample of typed supernovae in
which grain formation can be reliably tested. In addition, by studying
the evolution of recent supernovae in real time we can obtain
information about the physics of the dust-formation process.  In order
to disentangle the two possible origins of the NIR radiation, it is
vital to acquire IR observations spanning as wide a wavelength and
temporal range as possible. The longer NIR wavelengths ($HKLM$ bands)
are especially important since the flux at these wavelengths is less
likely to be contaminated by scattered light or strong H, He emission
lines, allowing a simple thermal interpretation. \\

\begin{figure*}
\vspace{15.5cm} \includegraphics{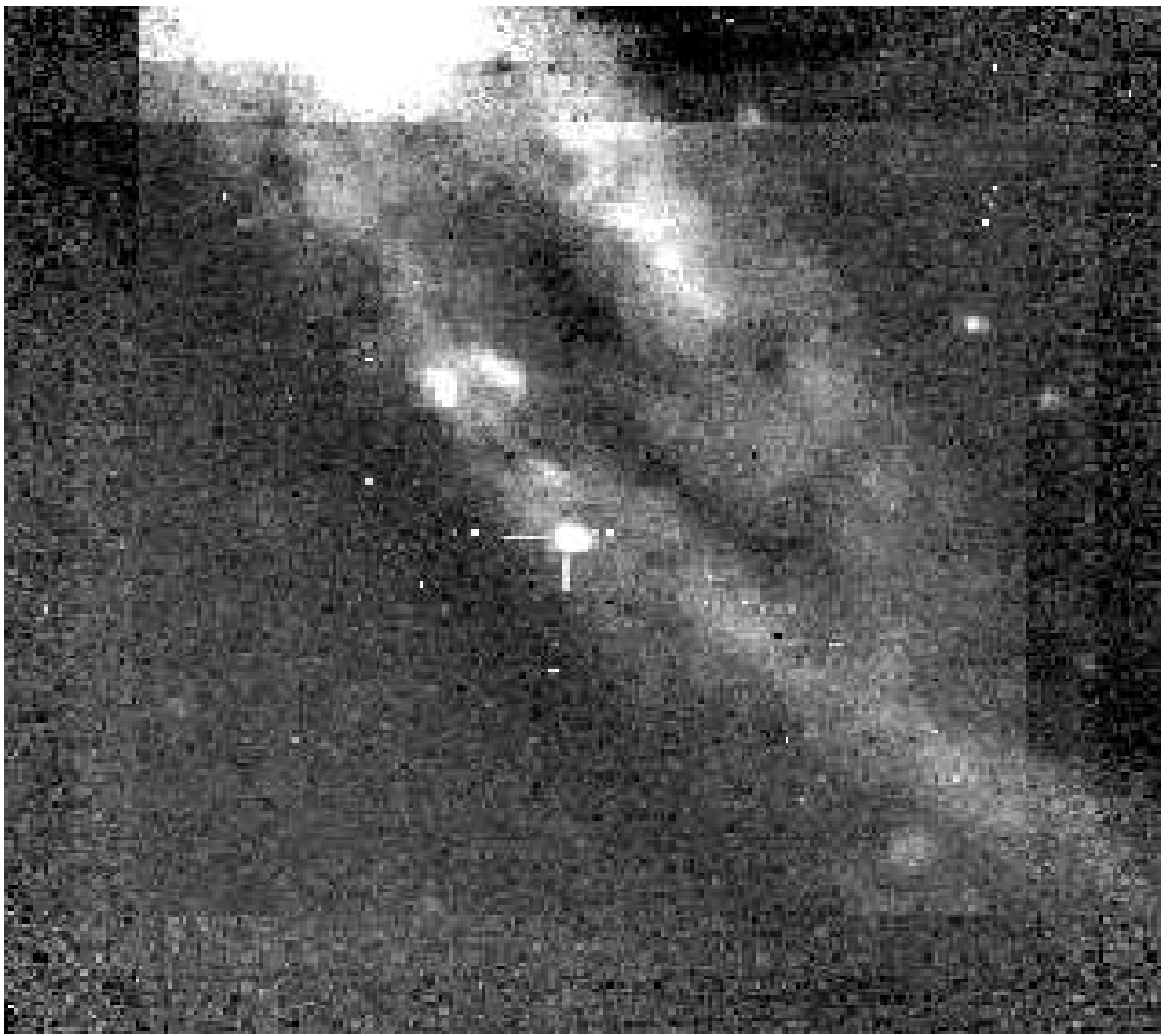}
\caption[SN~1998S] {A $J$-band image (83'' $\times$ 83'') showing SN~1998S (centre) at 333
days after shock breakout. North is to the top and east to the
left. The image was obtained with the United Kingdom Infrared Telescope, Hawaii.}
\label{sn98S}
\end{figure*}

An opportunity to address both the progenitor problem and the question
of supernovae as dust sources was provided by the occurrence of the SN
IIn 1998S.  This is the most intensively-studied type~IIn event (Bowen
et al. 2000; Fassia et al. 2000, 2001; Gerardy et al. 2000, 2002;
Leonard et al. 2000; Lentz et al. 2001; Liu et al. 2000; Roscherr \&
Schaefer 2000; Anupama, Sivarani \& Pandey 2001; Chugai 2001; Wang et
al. 2001; Chugai et al. 2002; Gruendl et al. 2002; Li et al. 2002;
Pooley et al. 2002; Meikle et al. 2003).  

\subsection{SN~1998S at early epochs}\label{early}

SN~1998S is one of the brightest SN IIn ever seen, reaching
$V=+12.17$ on 1998 March 18 (Fassia et al. 2000).  It was discovered
by Z. Wan (Li \& Wan 1998) on 1998 March 2.68 UT (JD 2,450,875.18) in
NGC 3877, an edge-on Sc spiral galaxy.  The supernova lies about 16''
west and 46'' south of the galaxy nucleus (see Fig.~1).  It was
identified as a SN~IIn by Filippenko \& Moran (1998).  Liu et
al. (2000) point out that the $BV$ light curves together with the
spectroscopic behaviour at 2--3 months are similar to those of the
type IIL SN~1979C.  Hamuy (2003) suggests that SNe IIn are
actually a subset of what he terms ``dense wind'' type~II supernovae
which includes SN~1979C.  A pre-discovery limiting magnitude of
$\sim$+18 obtained on 1998 February 23.7 (Leonard et al. 2000)
indicates that SN~1998S was discovered within a few days of shock
breakout.  Following Chugai (2001) we adopt 1998 February 24.7 UT = JD
2,450,869.2, 6~days prior to discovery, as the explosion epoch,
$t=0$~days.  Relative to this epoch, the zero epochs adopted by other
authors are later by, respectively, +5~days (Leonard et al. 2000),
+6~days (Fassia et al. 2000, 2001; Pooley et al. 2002), $\sim$+23~days
(Bowen et al. 2000; Gerardy et al. 2000, 2002).  Throughout this
paper, epochs quoted from other authors have been shifted to our zero
epoch definition.  We adopt a distance of 17~Mpc (Tully 1988), but
note that a smaller value of 15.5 Mpc is sometimes adopted (Sanders \&
Verheijen 1998; Gerardy et al. 2002).\\

The earliest optical spectra of SN~1998S show a blue continuum with
emission features superimposed. A rough blackbody fit yields
T$\sim$25,000~K, but with a blue excess (Leonard et al. 2000; Fassia
et al.  2001).  The emission lines are identified with H\,{\sc i}
(Balmer series), He\,{\sc i}, He\,{\sc ii}, C\,{\sc iii} and N\,{\sc
iii}.  The high-ionisation carbon and nitrogen lines are also commonly
observed in Wolf-Rayet stars (Leonard et al. 2000). The emission lines
have a broad base (e.g., H$\alpha$ FWZI$\sim$20,000 km s$^{-1}$), but
a narrow `peaked' unresolved centre.  The lines are symmetrical about
the local standard of rest.  This is quite surprising since at such an
early phase most of the receding part of the supernova should be
occulted by the photosphere.  However, this may actually constitute
some of the earliest evidence of a strong ejecta-CSM wind interaction.
It has been shown theoretically that, at this phase, one might expect
the photosphere to reside in a geometrically thin, opaque cool dense
shell (CDS) formed by the radiative shock propagating into the
extended stellar envelope (Grasberg, Imshennik \& Nadyozhin 1971; Falk
\& Arnett 1977; Chevalier \& Fransson 1985; 1994).  Chugai (2001) has
argued that these lines originate from the CDS and that the line
broadening is caused by the multiple Thomson scattering of line
photons on thermal electrons of the circumstellar gas.  The weakness
of broad ejecta lines during this early era is attributed to the sharp
photosphere at the SN ejecta boundary.  The blue excess can also be
attributed to the CDS, since the significant optical depth can yield
an increase in continuum absorptive opacity with wavelength, due to
both bound-free, (i.e., the Paschen continuum) and free-free processes
(Leonard et al. 2000). \\

By about 3~weeks after the explosion, the emission lines had almost
totally vanished.  This disappearance is attributed to the dense
inner-CSM being overrun by the ejecta.  This persisted for $\sim$3--7
weeks. During this time weak narrow lines are present, superimposed on
the continuum.  These are caused by the photoionization and heating of
the undisturbed wind by the supernova flash and subsequent X-ray
emission from the shocked wind and ejecta.  High-resolution echelle
spectra of these lines were obtained by Bowen et al. (2000) and Fassia
et al. (2001) on days~23 and 42.  These observations succeeded in
resolving the lines.  Between day~23 and day~42 the red wing of the
[O\,{\sc iii}]~5007~\AA\ profile moved redward, broadening the line,
so that by the latter epoch the profile was symmetrical about a
velocity of +847~km s$^{-1}$.  A high-resolution observation of
Gruendl et al. (2002) at 1 year showed that the [O\,{\sc
iii}]~5007~\AA\ profile was still symmetrical about this redshift.
This behaviour can be explained as being due to the effect of the
finite light travel time across a CSM having a denser outer zone.  As
the radiation-induced ionisation (due to the flash + shock radiation)
propagates across the CSM, it takes longer for the resulting nebular
emission to reach us from the far side.  At any given time, therefore,
the observer sees radiation from an ellipsoidal surface propagating
through the CSM (Morrison \& Sartori 1969).  On day~23, the vertex
region of the ellipsoid had not yet reached the inner boundary of the
denser zone, thus accounting for the weakness of the red wing of the
line.  However, the lack of any further shift in the line centre
between 46~days and 1~year indicates that the vertex had already
reached the dense zone by 46~days.  The early evolution of the line
asymmetry can be quantitatively explained if the inner boundary of the
denser zone was at a distance of 16.5 light days (2850~AU).  We note
that this distance is consistent with the observed persistence of the
line for at least a year since, assuming a maximum ejecta velocity of
10,000 km s$^{-1}$ (Leonard et al. 2000; Fassia et al. 2001), at
1~year it would still only have reached 2050~AU and so would not yet
have disturbed the dense outer CSM region (but see subsection 4.2
below).  Fassia et al. (2001) adopted +847~km s$^{-1}$ as the centre
of mass velocity for the SN~1998S system. Justification for this is
provided by the light travel time argument together with the lack of
any further shift between day~46 and 1~year.  From the narrow
forbidden lines such as [O\,{\sc iii}]~5007~\AA, an undisturbed CSM
velocity of about 40~km s$^{-1}$ is obtained, which is characteristic
of a red supergiant (RSG) wind.  From the intensity ratio of [O\,{\sc
iii}] (4959~\AA\ + 5007~\AA) to [O\,{\sc iii}] 4363~\AA, Fassia et
al. (2001) infer a CSM density of at least $1.5\times10^6$~cm$^{-3}$
at 185~AU, implying a CSM mass exceeding 0.005~M$_{\odot}$, and a
mass-loss rate exceeding around $2\times10^{-5}$~M$_{\odot}$
yr$^{-1}$.  This is consistent with the radio/X-ray estimate (scaled
to a 40~km~s$^{-1}$ wind velocity) of $\sim5\times10^{-4}$~M$_{\odot}$
yr$^{-1}$ (Pooley et al. 2002), and is comparable to that of SN 1979C
which had a mass loss rate of $\sim1.2\times10^{-4}$~M$_{\odot}$
yr$^{-1}$ (Lundqvist \& Fransson 1988) . \\

The behaviour of the narrow allowed H~{\sc i}, He~{\sc i} CSM lines
was more complex.  Not only did they exhibit asymmetric P~Cygni
profiles, but there were clearly two velocity components.  The slower
component is attributed to the same origin as the forbidden lines {\it
viz.} the photo-ionised, unaccelerated CSM.  The profile of this
component was probably a combination of emission from recombination
and possibly collisional excitation, together with a classical P~Cygni
line due to scattering from the populated excited levels.  The broad
absorption component is more difficult to explain.  It extends to a
velocity of around 350~km s$^{-1}$ which is too fast for a RSG wind.
It may be that, as in the case of SN~1987A, the SN~1998S progenitor
went through a fast-wind phase prior to explosion (Fassia et
al. 2001).  An alternative explanation is that the CSM close to the
supernova was accelerated by photospheric photons, or by relativistic
particles from the ejecta/CSM shock (Chugai et al. 2002).  Another
possibility is that the faster component arose in shocked clumps
within the CSM wind (Chugai et al. 2002).  By day~50, broad emission
lines in P${\beta}$ and He~{\sc i} 1.083~$\mu$m had appeared. The
earliest second season optical spectra (day~78) showed a similar
development in H${\alpha}$ and the Ca\,{\sc ii} triplet (Fassia et
al. 2001). The line profiles became increasingly ``square-shaped'',
characteristic of strong interaction between the expanding ejecta and
a dense CSM (Leonard et al. 2000; Fassia et al. 2000).  This was
confirmed by the detection of radio and X-ray emission (Pooley et
al. 2002) plus an unusually slow decline in V and I by $\sim$300 days
(Li et al. 2002).  \\

In summary, the wind of the SN~1998S progenitor underwent discrete
changes in the rate of mass outflow, resulting in several distinct CSM
zones.  The fading of the early-time broad emission components by
about day~18 indicates that the radius of the first boundary was less
than 100~AU (Leonard et al. 2000; Fassia et al. 2001). Beyond this
boundary the density presumably fell quite sharply.  The later
appearance of broad emission features led Leonard et al.  (2000) and
Fassia et al. (2001) to suggest the presence of a second boundary
where the CSM underwent a density increase.  For the explosion epoch
adopted here, the reappearance occurred between days~42 and 50
corresponding to $260\pm20$~AU.  However, the reality of this second
boundary was contested by Chugai et al. (2002) who suggested the
reappearance was due simply to the CDS becoming optically
thin. Finally, as argued above, the evolution of the narrow-line
profile suggests a third boundary at about 2850~AU where the CSM again
underwent a density increase.  For a wind velocity of 40~km s$^{-1}$,
these three boundaries correspond to material which left the
progenitor at epochs $-12$~yr, $-30$~yr and $-340$~yr relative to the
explosion date.  Pooley et al. (2002) also note that the best fit to
their radio data requires significant clumping or filamentation in the
CSM, a condition already inferred in two other type~II SNe, 1986J and
1988Z. Early-time spectropolarimetry by Leonard et al. (2000) and Wang
et al. (2001) indicate asphericity in the ejecta, CSM or possibly
both. \\

SN~1998S is unique in that it allowed the first-ever good IR
spectroscopic coverage of a SN~IIn (Gerardy et al. 2000; Fassia et
al. 2001).  In the $J$-band, we see Paschen $\beta$, Paschen $\gamma$
and He~{\sc i} 1.083~$\mu$m lines.  Their evolution was similar to
that seen in the optical.  At the earliest times broad-based, peaked
profiles were present. These faded by day~23, being replaced by broad,
square-shaped profiles by day~50.  Between days $\sim$15 and $\sim$60
a strong, unresolved He~{\sc i} 1.083~$\mu$m CSM line was superimposed
on the ejecta/CSM broad lines. By day~50, the $HK$-band was dominated
by Paschen~$\alpha$.  By day~114, first-overtone CO emission was
clearly present in the $K$-band.  The presence of CO in type~II SNe is
increasingly regarded as ubiquitous.  In all cases where $K$-band
observations have been carried out in the period 3-6 months
post-explosion, CO has been detected, e.g., SN~1987A (IIpec)
(Catchpole \& Glass, 1987; McGregor \& Hyland, 1987; Spyromilio et
al. 1988; Meikle et al. 1993), SN~1995ad (IIP) (Spyromilio \&
Leibundgut 1996), SN~1998S (IIn) (Gerardy et al. 2000; Fassia et
al. 2001), SN~1999dl (IIP) (Spyromilio, Leibundgut \& Gilmozzi 2001),
SN~1999em (IIP) (Spyromilio et al.  2001; Gerardy et al. 2002),
SN~1999gi (IIP) (Gerardy et al. 2002) and SN~2002hh (IIP) (Pozzo et
al., in preparation). In addition, CO has been detected in the type~Ic
SN~2000ew (Gerardy et al. 2002).  The rotation-vibration states of CO
are a powerful coolant (especially the fundamental band lines).  The
presence of CO is suspected to be a necessary condition for dust
condensation to occur in the ejecta.  Modelling of the SN~1998S
spectra suggests a CO velocity of $\sim$2000 km s$^{-1}$ (Gerardy et
al. 2000; Fassia et al. 2001).  From this, Fassia et al. (2001)
deduced a core mass of 4~M$_{\odot}$ implying a massive progenitor.
The actual mass of CO derived was 10$^{-3}$~M$_{\odot}$.  This is much
smaller than the $>0.1$~M$_{\odot}$ of C and $>1.0$~M$_{\odot}$ of O
likely to be present in the ejecta (Woosley \& Weaver 1995), implying
that most of the C and O remain in the gas phase and are thus
available to condense into, respectively, graphite and silicate
grains. \\

On day~136, Fassia et al. (2000) measured the IR flux out to a
wavelength of 3.8~$\mu$m ($L'$-band).  This revealed a remarkable IR
excess of $K-L'=+2.5$.  To produce the observed flux the lowest
possible radius of the IR emission region is given by a blackbody with
a temperature close to the dust evaporation temperature of
$\sim$1500~K.  For dust condensing in the ejecta to attain a
sufficiently large blackbody radius by 136~days would require an
expansion velocity of 11,000~km s$^{-1}$ (see Fassia et al. 2000).
Such high velocities were seen only in the extreme outer zones of the
H/He envelope. No metals were seen at such high velocities.  Fassia et
al. (2000) concluded that the IR excess at this epoch cannot,
therefore, have been due to grain condensation in the ejecta.  It must
instead have been produced by an IR echo of the maximum-light
luminosity from pre-existing dust in the CSM. This conclusion is based
only on the intensity of the $L'$-band flux.  It is strengthened if we
fit a blackbody to both the $K$ and $L'$ fluxes. After correction for
the optical photosphere, this yields $K-L'\approx3.7$ corresponding to
a temperature of only 650~K, requiring the dust to lie at an
equivalent velocity of at least 67,000~km s$^{-1}$! \\
 
SN~1998S was exceptionally luminous, reaching a de-reddened M$_B =
-19.6$ (Fassia et al. 2000). This is around $\times$10 the typical
luminosity of a type~II SN.  The excellent early-time coverage
achieved in the optical and NIR allowed Fassia et al. (2000) to examine
the bolometric light curve.  Both blackbody and UVOIR (i.e.,
ultraviolet-optical-infrared range) fits indicate that the total
energy radiated in the first 40~days exceeded $10^{50}$ ergs, which is
again $\times$10 the typical value for type~II SNe.  Between days 96
and 136, the bolometric light curve is well-reproduced by the
radioactive decay luminosity of 0.15~M$_{\odot}$ $^{56}$Ni.  However,
by this era the ejecta/CSM shock energy must also have been making
a contribution.

\subsection{SN~1998S at later epochs}

Much of the early-time behaviour of SN~1998S can be attributed to the
interaction of the supernova with a pre-existing, dusty, perhaps
disk-like CSM.  SN~1998S remained observable from X-rays to radio for
over 3 years (Pooley et al. 2002; this work) due to the ongoing
conversion of the SN kinetic energy to radiation via the ejecta/CSM
interaction. We continued regular observations during this phase, to
day~1242.  A brief, preliminary presentation of this work was given in
Meikle et al. (2003). Other late-time studies have been presented:
optical spectroscopy to 499~d (Leonard et al. 2000); high-resolution
spectroscopy at $\sim 370$~d (Gruendl et al. 2002); NIR spectroscopy
to $\sim 380$~d and NIR photometry to 819~d (Gerardy et al. 2000,
2002); X-ray observations to 1055~d and radio observations to 1065~d
(Pooley et al. 2002).\\

In this paper we present and discuss NIR photometry of SN~1998S
spanning 333 to 1242 days post-explosion, optical spectra covering 305
to 1093~days and IR spectra covering 333 to 1191~days.  The NIR
photometry presented here extends as far as the $M'$-band (CW
4.7~$\mu$m), the first time that such a long IR wavelength has been
detected in any SN other than the exceptionally close SN~1987A.  The
paper is organised as follows. NIR photometry and optical/NIR
spectroscopy are presented in Section~2.  The evolution of the
H$\alpha$ and He~{\sc i}~1.083~$\mu$m profiles are examined in detail,
and for this purpose the data are augmented by spectra from other
sources.  In Section~3 we compare the NIR spectral energy distribution
(SED) with pure blackbody functions and blackbodies weighted by
emissivities $\lambda^{-1}$ and $\lambda^{-2}$.  In Section~4 we
examine possible energy sources for the post-300~d NIR emission, and
discuss the location of the IR-emitting dust. Evidence of episodic
mass-loss from the progenitor is discussed.  We also compare the
late-time $(K-L')_0$ colour evolution of SN~1998S with that of other
type~II SNe and suggest that it provides a useful identifier of
supernovae whose progenitors had a massive CSM.  Conclusions follow in
Section~5.

\begin{table*}
\centering
\caption{Log of infrared imaging of SN~1998S at UKIRT} 
\begin{minipage}{\linewidth}
\renewcommand{\thefootnote}{\thempfootnote}
\renewcommand{\tabcolsep}{0.4cm}
\begin{tabular}{clrll}
\hline
JD(2450000+) & Date &  Epoch\footnote{Days after explosion, assumed to
be 1998 February 24.7 UT (JD 2450869.2).} & Filters\footnote{See subsection 2.1
for details.} & Standard star\footnote{FS standards were used
for $JHK$, and HD standards for $L'M'$.} \\ \hline
1202.2  & 1999 Jan 23  &   333.0~   & $J$, $H$, $K$, $L'$                &  FS21, HD105601 \\
1243.2  & 1999 Mar  3  &   374.0~   & $J$, $H$, $L'$, nb$M$              &  FS15, HD84800, HD105601, HD106965 \\
1274.8  & 1999 Apr  4  &   405.5~   & $J$98, $H$98, $K$98              &  FS21  \\
1333.5  & 1999 Jun  2  &   464.2~   & $J$, $H$, $K$, $L'$, nb$M$           &  FS21, HD106965 \\
1533.2  & 1999 Dec 19  &   664.0~   & $J$98, $H$98, $K$98, $L'$98, $M'$98  &  FS130, HD105601 \\
1567.5  & 2000 Jan 23  &   698.2~   & $J$98, $H$98, $K$98, $L'$98, $M'$98  &  FS21, HD105601  \\
1579.0  & 2000 Feb  4  &   709.8~   & $M'$98                       &  HD105601 \\
1696.2  & 2000 May 30  &   827.0~   & $J$98, $H$98, $K$98, $L'$98, $M'$98  &  FS21, HD105601 \\
1916.8  & 2001 Jan  7  &  1047.5~  & $J$98, $H$98, $K$98, $L'$98, $M'$98  &  FS21, HD105601, HD106965 \\
2067.0  & 2001 Jun  5  &  1197.8~  & $J$98, $H$98, $K$98              &  FS131 \\
2111.5  & 2001 Jul 19  &  1242.2~  & $L'$98                       &  HD106965 \\
\hline
\vspace{-0.8cm}
\end{tabular}
\end{minipage}
\end{table*}

\begin{table*}
\centering
\caption[]{Infrared photometry of SN~1998S}
\begin{minipage}{\linewidth}
\renewcommand{\thefootnote}{\thempfootnote}
\renewcommand{\tabcolsep}{0.5cm}
\begin{tabular}{crccccc} \hline
 JD(2450000+) &  Epoch\footnote{Days after explosion.}& $J$ & $H$ & $K$ & $L'$ & $M'$ \\ \hline
 1202.2  &     333.0  &   ~16.653(12)\footnote{Figures in brackets give the statistical error, in units of the magnitude's least significant one or two digits.} & 14.809(5)~~ & 13.344(2)~ & 11.560(28) &  -- \\  
 1243.2  &     374.0  &   16.799(34)   &  14.922(13)  &   -- &   11.52(17)~ &   11.291(86) \\
 1274.8  &     405.5  &   17.077(25)   &  15.269(8)~  &   13.431(2)~ &      --      &      --      \\
 1333.5  &     464.2  &   17.609(29)   &  15.537(12)  &   13.806(4)~ &   11.69(13)~ &  11.14(16)~  \\
 1533.2  &     664.0  &   18.025(33)   &  16.421(14)  &   14.448(4)~ &   11.856(18) &   11.44(14)~ \\
 1567.5  &     698.2  &   18.341(35)   &  16.728(15)  &   14.556(4)~ &   11.970(26) &   11.82(14)~ \\
 1579.0  &     709.8  &      --        &     --       &      --      &       --     &   12.30(12)~ \\
 1696.2  &     827.0  &   18.680(27)   &  17.148(16)  &   14.992(4)~ &   12.200(33) &      $>11.9(2\sigma)$\footnote{For this epoch we did not detect the SN in the M' filter band image.}      \\
 1916.8  &    1047.5  &   19.280(48)   &  17.956(27)  &   15.843(7)~ &   12.985(39) &   12.84(15)~ \\
 2067.0  &    1197.8  &   19.798(83)   &  18.439(77)  &   16.366(13) &   [13.29(3)]\footnote{Estimated by linear interpolation within $L'$-band light curve.}      &      --      \\
 2111.5  &    1242.2  &      --        &     --       &   [16.52(2)]\footnote{Estimated by linear extrapolation of $K$-band light curve.} &   13.40(15)~ &      --      \\ \hline
\vspace{-0.8cm}
\end{tabular} 
\end{minipage}
\end{table*}

\section{Observations}

\subsection{Near-infrared photometry}\label{nirphot}

$JHKL'M'$ images of SN~1998S were obtained in the period 333-1242d at
the 3.8m United Kingdom Infrared Telescope (UKIRT) on Mauna Kea
(Hawaii).  The NIR observations for the earlier epochs were obtained
with the IRCAM3 camera. This contains a 256$\times$256 InSb array,
originally with a 0.286 arcsec/pixel plate scale. In late 1999, the
camera was reconfigured to a plate scale of 0.081 arcsec/pixel, and
renamed TUFTI. Also during 1999, a new camera, UFTI, was introduced.
This contains a 1024$\times$1024 HgCdTe array with a 0.091
arcsec/pixel scale.  Most of the later $JHK$ images were obtained with
UFTI, with the $L'$ and $M'$ images being taken with TUFTI (see
Table~1 for details).  IRCAM3 used the older $J$(Barr), $H$(Barr),
$K$(OCLI), $L'$(OCLI) and $nbM$ filters at effective central
wavelengths of 1.25, 1.65, 2.205, 3.8 and 4.675 $\mu$m; UFTI/TUFTI
uses the newer Mauna Kea Filter set $J$98(MK), $H$98(MK), $K$98(MK),
$L'$98(MK) and $M'$98(MK) at effective wavelengths of 1.25, 1.635,
2.20, 3.77 and 4.68 $\mu$m (see Tokunaga, Simons \& Vacca 2002).  For
$JHK$, UKIRT faint standard stars were observed, while for $L'M'$ we
used HD standards.  Table~1 lists the log of the observations plus
information on the different filter sets and UKIRT standards.  5-point
and 9-point dither patterns were used in the $JHK$ and $L'M'$ filter
bands respectively. \\

\begin{table*}
 \centering
 \begin{minipage}{200mm}
 \caption{Infrared colours of SN~1998S}
 \renewcommand{\tabcolsep}{0.5cm}
 \begin{tabular}{@{}ccccccc@{}}
 \hline
 JD(2450000+) & Epoch (days) &  $J-H$ & $H-K$         & $K-L'$     &  $L'-M'$ \\ \hline
 1202.2  &     333.0  & 1.844(13)\footnote{Figures in brackets give the statistical error, in units of the magnitude's least significant one or two digits.}~  & 1.465(5)~   & 1.784(28)~  &  -- \\
 1243.2  &     374.0  & 1.877(36)~  &    --       &    --       & 0.23(19) \\
 1274.8  &     405.5  & 1.808(26)~  & 1.838(8)~   &    --       &  -- \\
 1333.5  &     464.2  & 2.072(31)~  & 1.731(13)   & 2.11(13)    & 0.55(20) \\
 1533.2  &     664.0  & 1.604(36)~  & 1.973(15)   & 2.592(18)~  & 0.41(14) \\
 1567.5  &     698.2  & 1.613(38)~  & 2.172(16)   & 2.586(26)~  & 0.15(14) \\
 1696.2  &     827.0  & 1.532(31)~  & 2.156(16)   & 2.792(33)~  & -- \\
 1916.8  &    1047.5  & 1.324(55)~  & 2.113(28)   & 2.858(40)~  & 0.15(15) \\
 2067.0  &    1197.8  & 1.36(11)    & 2.073(78)   & [3.08(3)]\footnote{Estimated by linear interpolation within $L'$-band light curve.}   --  & -- \\
 2111.5  &    1242.2  &    --       &      --     & [3.12(15)]\footnote{Estimated by linear extrapolation of $K$-band light curve.}   --  & -- \\ \hline
\vspace{-0.8cm}
\end{tabular} 
\end{minipage}
\end{table*}

The images were reduced using the standard Starlink packages {\sc ircamdr}
(Aspin 1996) and {\sc orac-dr} (Economou et al. 2003).  Magnitudes were
measured via aperture photometry within the Starlink package
{\sc gaia}\footnote{Graphical Astronomy and Image Analysis Tool, version
2.6-9} (Draper, Gray \& Berry 2002). The sky background was measured
using a concentric annular aperture. Choice of target aperture and sky
annulus was a compromise between maximising the S/N, and minimising
the effects of the galaxy background gradients and structure.  An
aperture radius of 1.4 arcsec was selected, this being equivalent to 5
pixels for IRCAM3, 15.4 pixels for UFTI and 17.3 pixels for TUFTI.
The annulus was set to cover $\times$3 the area of the target aperture
in order to minimise the effect of statistical uncertainty in the
background estimation (Merline \& Howell 1995).  Thus, the annulus was
chosen to have inner and outer radii, respectively, $\times$1.5 and
$\times$2.5 that of the aperture.  Statistical uncertainty in the
photometry was determined by the sky variance method, with 2 sigma
clipping rejection (all points in the sky background measurement which
deviate by more than 2 standard deviations from the mean are
rejected).  Magnitudes were determined by comparison with the standard
stars listed in Table~1.  Systematic uncertainties arising from
varying atmospheric conditions from epoch to epoch or between the
target and standard are discussed in Section 3. \\ 

\begin{figure}
\vspace{7.3cm} \includegraphics{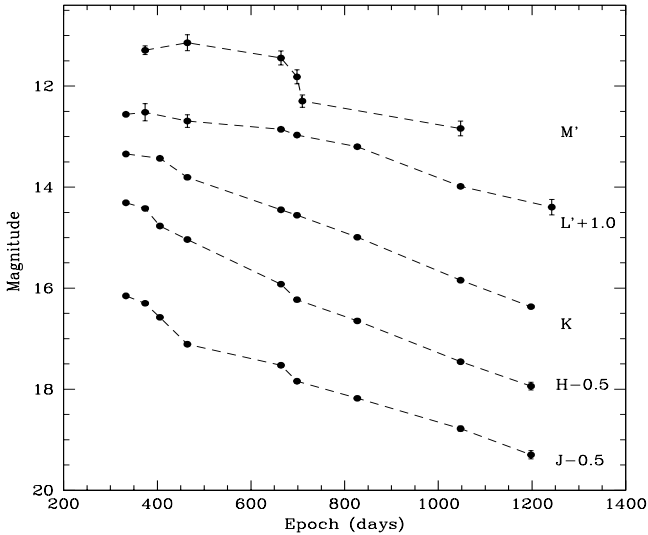}
\caption[]{Post-300~d $JHKL'M'$ light curves of SN~1998S. Only the
statistical errors are shown (see text).}
\label{lightc_jhklm}
\end{figure}

\begin{figure}
\vspace{7.3cm}\includegraphics{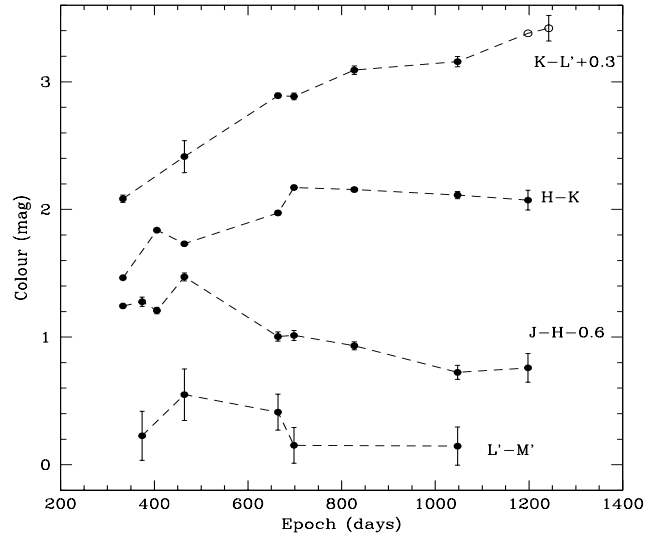}
\caption[]{Post-300~d IR colour evolution of SN~1998S. The latest
two $K-L'$ values (open circles) were estimated by linear
interpolation between $L'$-band values (penultimate point) or linear
extrapolation of the $K$-band values (final point).  Only the
statistical errors are shown (see text).}
\label{lightc_jhklm_col_plus_extrap}
\end{figure}

The $JHKL'M'$ magnitudes for SN~1998S are listed in Table~2 together
with associated statistical errors in parentheses. The quoted errors
do not include additional uncertainty due to differences in conditions
under which the supernova and standards were observed.  The light
curves are shown in Fig.~2.  They extend to longer wavelengths and to
later phases than has ever been achieved for the IR light curves of
any supernova other than SN~1987A.  A generally monotonic decline is
seen in the $JHKL'$ bands.  The $M'$-band light curve shows a sudden
fading of about 1~mag. at around 700~days, although the errrors are
quite large.

The NIR colours for each epoch are listed in Table~3, and their
evolution is illustrated in Fig.~3.  We note in particular the 
increase in $H-K$ to day$\sim$700~d and in $K-L'$ to beyond
1000~d. The latter reached a value of 2.9 by 1047~d.  Indeed, using
modest linear extrapolation of the $K$-band magnitudes we see that by
1242~d, $K-L'$ has reached a value of 3.1.

\subsection{Spectroscopy}

We obtained optical spectroscopy at the Nordic Optical
Telescope (NOT) and the Isaac Newton Telescope (INT), both on La
Palma.  Infrared spectroscopy was obtained at the United Kingdom
Infrared Telescope (UKIRT), Hawaii. Tables~4 and 5 give the observing
logs of the optical and IR observations respectively. For the IR
spectroscopy, the standard BS4431 was used.\\

\begin{table*}
\centering
\caption{Log of optical spectroscopy of SN~1998S}
\begin{minipage}{\linewidth}
\renewcommand{\thefootnote}{\thempfootnote}
\renewcommand{\tabcolsep}{0.35cm}
\begin{tabular}{ccclcccc} 
\hline 
JD\footnote{2450000+}  & Date       & Epoch  & Telescope/ & Spectral & Spectral& Slit width & Spectrophotometric \\ 
       & (UT)       &  (d)   & Instrument & range    & res. &  (arcsec)  & standard  \\
          &            &        &            & (\AA)    &  (\AA)  &            &           \\ \hline
1174.5 & 1998 Dec. 27  & 305.3   & NOT/ALFOSC & 5825-8344 & 3.0     &   1.2    & G191b2b        \\
1215.6 & 1999 Feb. 6.1 & 346.4  & INT/IDS    & 4350-7700 & 5.0      &   1.2      & Feige 34  \\
1287.5 & 1999 Apr. 19  & 418.3   & NOT/ALFOSC & 5100-9095 & 8.4    &   1.3     & Feige 34  \\ 
1343.5 & 1999 Jun. 14  & 474.3  & NOT/ALFOSC & 5804-8322 & 3.0    &   1.3      & Feige 56  \\
1962.6 & 2001 Feb. 22.1 & 1093.4& NOT/ALFOSC & 4000-8500 & 8.3    &   1.3      & Feige 34  \\
``     &  ``            & ``    & ``         & 5802-8340 & 3.0    &    ``      &  ``       \\ \hline
\\ 
\vspace{-0.8cm}
\end{tabular} 
\end{minipage}
\end{table*}

\begin{figure*}
\vspace{6cm} 
\includegraphics{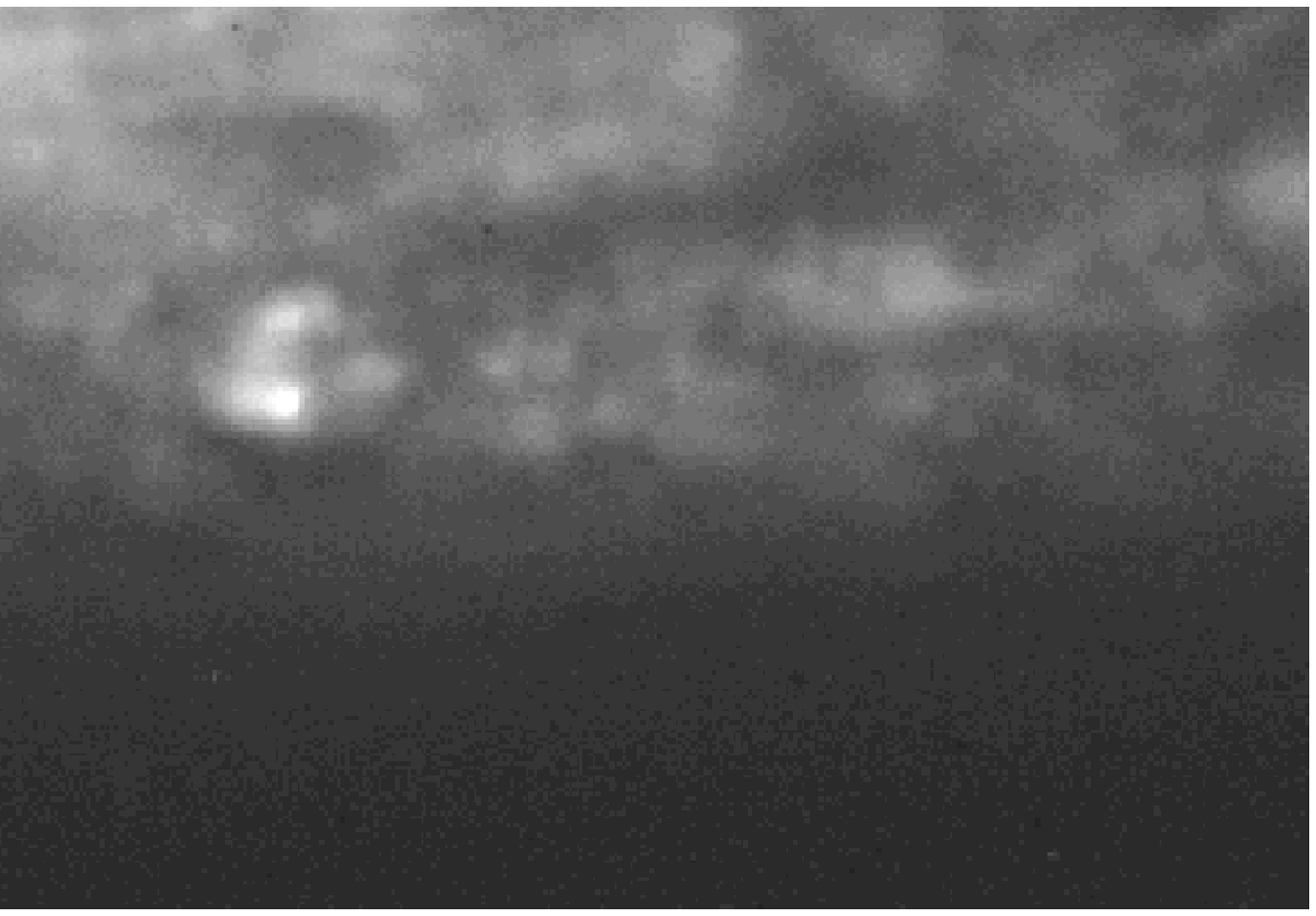}
\includegraphics{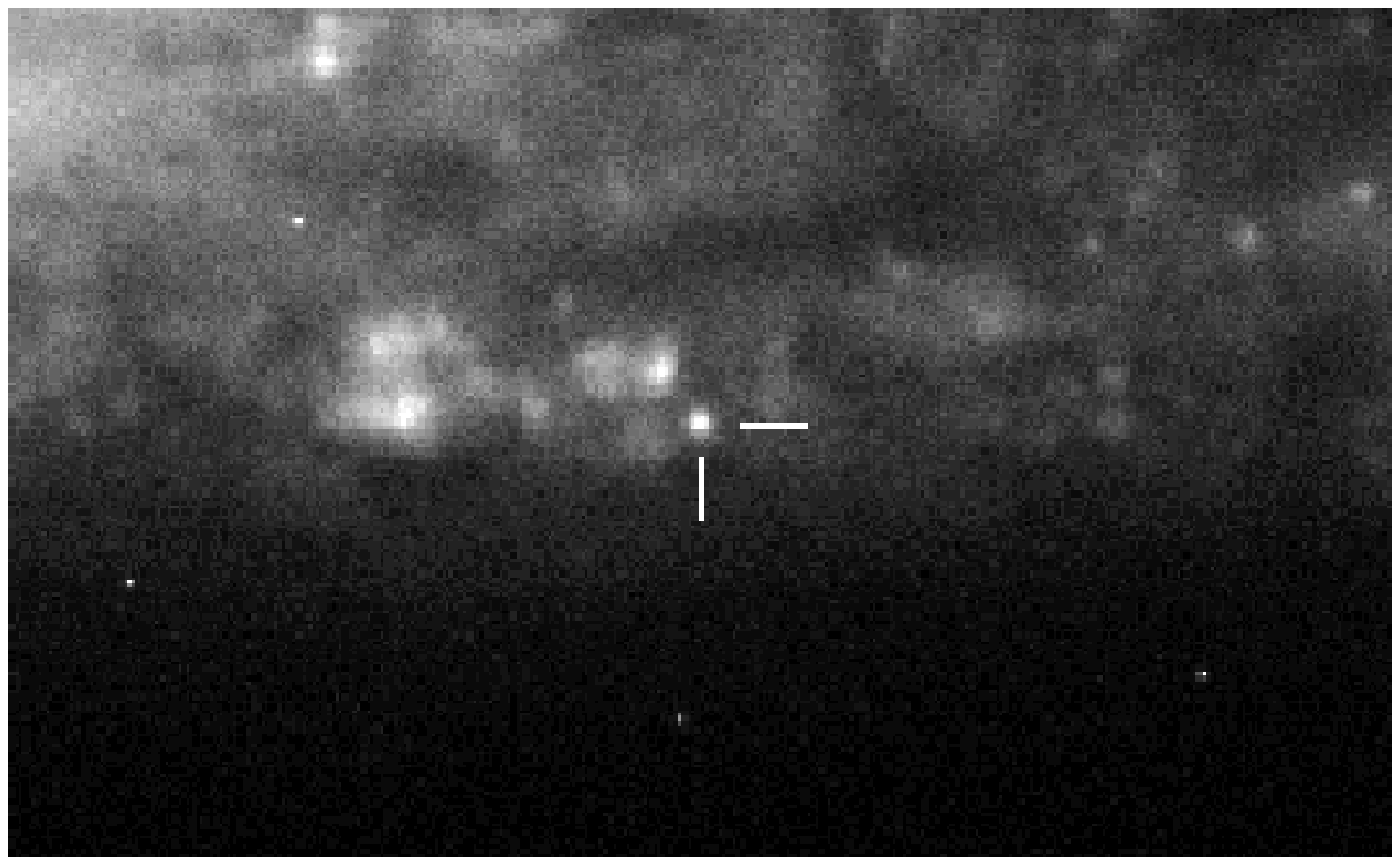}
\caption[]{$V$-band (left hand image) and H$\alpha$ (right hand image)
images of SN~1998S taken with ALFOSC at the Nordic Optical Telescope
at 1093~days post-explosion.  North is approximately towards the top
left-hand corner.  The images cover 69 arcsec. $\times$ 42 arcsec. }
\end{figure*}

The day~1093 NOT spectra, the INT spectrum and the UKIRT spectra were
reduced using the standard routines of the {\sc figaro} package (Shortridge
1991). The other NOT spectra were reduced using the standard procedures
within {\sc iraf}.\footnote{{\sc iraf}(Image Reduction and Analysis Facility) 
is distributed by the National Optical Astronomy Observatory (NOAO), which
is operated by the Association of Universities for Research in
Astronomy (AURA), Inc. under cooperative agreement with the National
Science Foundation.}  The fluxing of the day~1093 NOT spectrum was corrected 
using contemporary photometry in a narrow band H$\alpha$ filter 
(CW 6556~\AA, width 162~\AA). The H$\alpha$ image is shown in Fig.~4 
together with a contemporary $V$-band image.  The source of the H$\alpha$ 
emission is clearly visible as a point source in the H$\alpha$ image. 
From these we obtain approximate magnitude estimates of: $V=21.25\pm0.6$ and
$m_{H\alpha}=20.15\pm0.3$.  A contemporary $R$-band image yields
$R=20.3\pm0.5$.  The IR spectra fluxing was adjusted to match the IR
photometry in Table~2. The spectra are plotted in Figs. 5, 6
(optical) and 7 (IR).\\

\begin{table}
\centering
\caption{Log of UKIRT/CGS4 infrared spectroscopy of SN 1998S}
\begin{minipage}{\linewidth}
\renewcommand{\thefootnote}{\thempfootnote}
\renewcommand{\tabcolsep}{0.09cm}
\begin{tabular}{clcccc} 
\hline 
JD\footnote{2450000+}  & Date & Epoch  & Spectral & Spectral & Slit width \\ 
 & (UT) &(d)&range  &    res.  & (arcsec) \\
& &  &  ($\mu$m) &  ($10^{-4} \mu$m) \\ \hline
1203.0 & 1999 Jan. 24.5 & 333.8 & 1.01-1.33 & 12.5 & 0.61 \\
       &                &       & 1.45-2.51 & 25  & 0.61 \\ 
1228.1 & 1999 Feb. 18.6 & 358.9 & 1.01-1.33 & 25  & 1.23 \\
1273.1 & 1999 Apr.  4.6 & 403.9 & 1.01-1.33 & 25  & 1.23 \\ 
1567.2 & 2000 Jan. 23.7 & 698.0 & 1.02-1.34 & 25  & 1.23 \\
1578.1 & 2000 Feb.  3.6 & 708.9 & 1.02-1.34 & 25  & 1.23 \\
   &                &           & 1.84-2.48 & 50   & 1.23 \\
1918.2 & 2001 Jan.  8.7 & 1049.0 & 1.79-2.40 & 50  & 2.46 \\
2060.8 & 2001 May  31.3 & 1191.6 & 1.03-1.35 & 25 & 1.23 \\ 
   &                &            & 1.83-2.46 & 50  & 1.23 \\ \hline
\vspace{-0.8cm}
\end{tabular} 
\end{minipage}
\end{table}

The optical spectra are dominated throughout by a broad, complex
H$\alpha$ profile. This is described in more detail below. At a much
lower level, however, broad emission from a few other species can be
identified (Fig.~6). Between days 305 and 474, we identify He\,{\sc i}
5876~\AA\ and [Ca\,{\sc ii}] 7291, 7324~\AA.  On day~418 the spectrum
extends far enough to the red to reveal a blend of [O\,{\sc
i}]~8446~\AA\ and [Ca\,{\sc ii}] 8498,~8542,~8662~\AA. The He\,{\sc i}
feature exhibits a peak blueshifted by $\sim$--4000~km s$^{-1}$, as
does the [O\,{\sc i}]~8446~\AA\ peak (assuming it dominates the blend
with the calcium triplet).  These blueshifts are similar to those seen
in the H$\alpha$ line peaks (see below). There is some evidence of a
similarly blueshifted H$\beta$ peak in the day~346 spectrum. In the
same spectrum there may also be a [O\,{\sc iii}]~5007~\AA\ peak
blueshifted by about 3000~km s$^{-1}$.  The [Ca\,{\sc ii}] 7291,
7324~\AA\ line peak on days~305 and 418 shows a blueshift of only
about 2300~km s$^{-1}$. Given the larger blueshifts seen in the other
oxygen lines, this argues against a significant blended contribution
from [O\,{\sc ii}]~7319, 7330~\AA.  The broad feature lying at
5100-5400~\AA\ has been identified as an Fe~{\sc ii} emission band
(Garnavich, Challis \& Kirshner 1998).  On day~1093, we identify H$\beta$, 
He\,{\sc i} 5876~\AA, [O\,{\sc i}]~6300~\AA\ and [O\,{\sc iii}]~4363, 5007~\AA.  
A fainter feature at about 7275~\AA\ could be due to [O\,{\sc ii}]~7319, 
7330~\AA, possibly blended with [Ca\,{\sc ii}] 7291, 7324~\AA.  The profiles 
appear to have a similar form to that of H$\alpha$ at this epoch, with much of
the flux appearing as a broad feature to the blue of the local rest
frame wavelengths.  Narrow peaks at the rest wavelengths are also
seen, although these may be simply unresolved emission from a
coincident H~II region. However, we note that a higher resolution
spectrum of the much stronger H$\alpha$ feature (Fig.~8) reveals a
rest-frame peak with wings extending to $\pm$1000~km s$^{-1}$, making
a H~II region origin unlikely, at least for this line (see below). \\

\begin{figure}
\vspace{10cm} \includegraphics{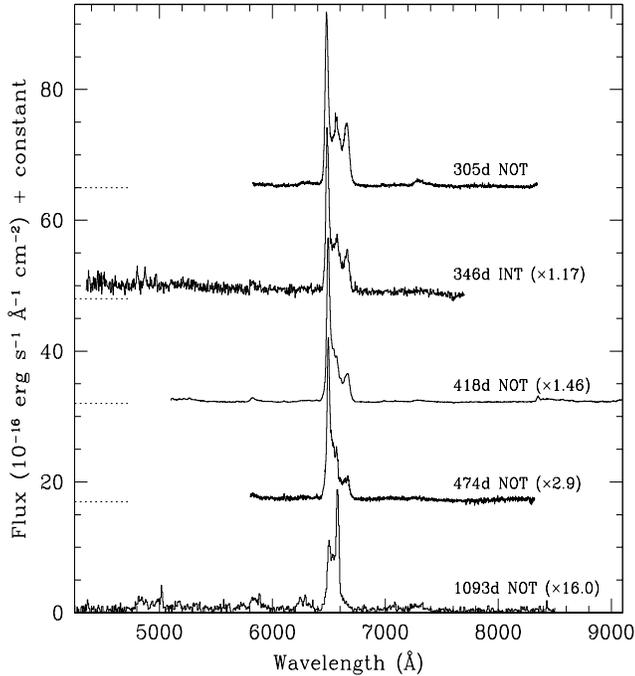}
\caption[]{Post-300~d optical spectra of SN~1998S obtained with the
IDS spectrograph at the Isaac Newton Telescope or with the ALFOSC
spectrograph on the Nordic Optical Telescope, both on La Palma.  The
zero flux levels are shown by the horizontal dotted lines on the
left-hand side. In addition, the spectral fluxes have been scaled by
the factors shown in brackets. Throughout the period covered, the
spectrum is dominated by H$\alpha$ emission. For more detailed
displays of these spectra see Figs.~6 (weaker lines) and 8 (H$\alpha$
lines).}
\end{figure}

\begin{figure}
\vspace{10cm} \includegraphics{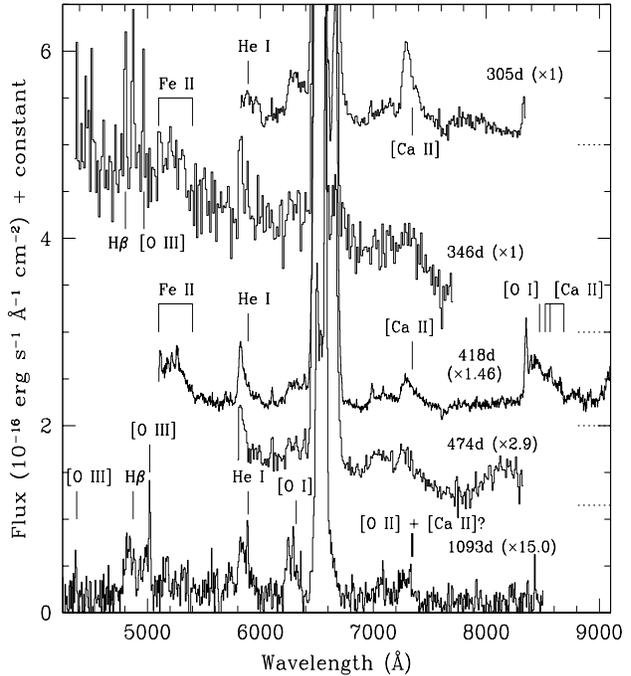}
\caption[]{Post-300~d optical spectra of SN~1998S, plotted to show the
weaker features. The line wavelengths in the SN~1998S restframe are
usually indicated. The exception is in the day~346 spectrum where, for
clarity, the blueshifted peaks of H$\beta$ and [O\,{\sc
iii}]~5007~\AA\ are shown.  The zero flux levels are shown by the
horizontal dotted lines on the right-hand side. In addition, the
spectral fluxes have been scaled by the factors shown in brackets.}
\end{figure}

\begin{figure}
\vspace{10cm} \includegraphics{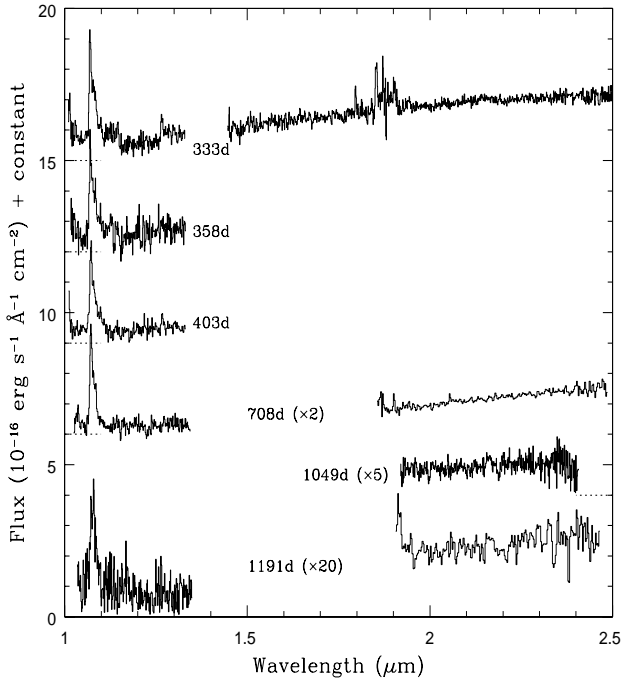}
\caption[]{Post-300~d IR spectra of SN~1998S obtained with the CGS4
spectrograph on UKIRT, Hawaii. The spectra have been displaced
vertically for clarity. The zero flux levels are shown by the
horizontal dotted lines on the left-hand side or, for 1049~d, on the
right. In addition, the spectral fluxes for the latest three epochs
have been scaled by the factors shown in brackets.  For a more
detailed display of the He~I 1.083~$\mu$m profiles see Fig.~9}
\end{figure}

\begin{figure}
\vspace{10cm} \includegraphics{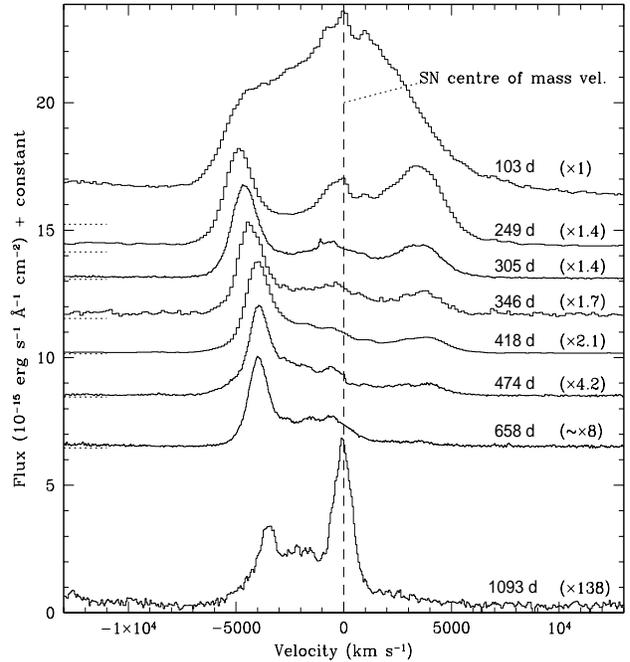}
\caption[]{Evolution of the H$\alpha$ profile, in velocity space. Zero
velocity corresponds to the rest frame of the SN~1998S system (see
text). The spectra have been scaled by the amounts shown in brackets,
in order to approximately normalise to the height of the blue-shifted
peak. They have also been displaced vertically for clarity.  The zero
flux levels are shown by the horizontal dotted lines on the left-hand
side.  The day~103 spectrum is from Fassia et al.  (2001), the day~249
spectrum from Gerardy et al. (2000), and the day~658 spectrum from
D. Leonard and T. Matheson (private communication). The day~658
spectrum was placed on an absolute flux scale by interpolation of the
H$\alpha$ intensity light curve.}
\end{figure}

The IR spectra are characterised by strong, broad complex
He\,{\sc i}~1.083~$\mu$m emission plus, when observed, a smooth continuum
rising to longer wavelengths. P$\alpha$ and P$\beta$ are also visible
in the day 333~spectrum.\\

\subsubsection{H${\alpha}$ and He\,{\sc i} 1.083~$\mu$m profiles}

In Fig.~8 we show the evolution of the H$\alpha$ profile between
days~103 and 1093.  To maximise the temporal coverage, we have added
spectra at +103~d (Fassia et al. 2001), +249~d (Gerardy et al. 2000)
and at +658~d (D. Leonard and T. Matheson, private communication).
The +658~d spectrum was placed on an absolute flux scale by
interpolation of the H$\alpha$ intensity light curve which declined at
about 1mag/155~days between days~249 and 1093. The velocities of the
peaks and extreme edges of the profile with respect to the SN
centre-of-mass velocity are listed in Table~6, together with the mean
flux and intensity of the line.

On day~103 the profile has the form of a broad, steep-sided, fairly
symmetrical line spanning $\pm$7000~km s$^{-1}$ across the base.  This
appearance persisted to at least day~145 (Leonard et al. 2000).
However, by the time the supernova was recovered in the second season,
the shape was remarkably different. The day~249 spectrum of Gerardy et
al. (2000) shows that the profile had developed a triple-peak
structure, comprising a central peak close to the rest-frame velocity,
and two outlying peaks at, respectively, --4860~km s$^{-1}$ and
+3400~km s$^{-1}$.  Gerardy et al. suggest that the outermost peaks
could have been produced by an emission zone having a ring or disk
structure seen nearly edge-on, and resulting from the SN shock-wave
collision with the disk/ring.  Following Chugai \& Danziger (1994)
they also suggest that the central peak might have been due to shocked
wind clouds.  Between days~249 and 658 we see a steady fading of the
central and red-shifted peaks with respect to the blue-shifted peak.
Gerardy et al. (2000) and Leonard et al. (2000) also observed this
effect, suggesting that dust condensation in the ejecta is
responsible. We note also that the strong blue peak shifted from
--4860~km s$^{-1}$ on day~249 to --3900~km s$^{-1}$ by day~474,
presumably due to a slowing of the shock as it encountered an
increasing mass of CSM. The velocity then remained fairly constant for
the subsequent $\sim$200~d.  Finally, when we recovered the supernova
in the 4th season on day~1093, it can be seen that the profile had
undergone another dramatic change.  While the blue-shifted peak
persisted, slowing slightly to --3430~km s$^{-1}$, the central peak
had grown in relative strength to about twice the height of the blue
peak. However, the total intensity of the day~1093 line is less than
10\% of the day~658 spectrum and so it is possible that the day~1093
central peak was present at earlier times but was swamped by other
stronger emission. \\

In Fig.~9 we show a sequence of He\,{\sc i} 1.083~$\mu$m spectra
obtained at UKIRT. Again to increase the temporal coverage we include
spectra obtained by Fassia et al. (2001) on day~114, and by Gerardy et
al. (2000) on days~247 and 283.  The velocities of the blue peak and
extreme edges of the profile with respect to the SN centre-of-mass
velocity are listed in Table~7, together with the mean flux and
intensity of the line.  Judging from the strength of P$\beta$
emission, the earlier epochs were probably slightly contaminated by
P$\gamma$ emission.  Consequently the velocities of the red edges of
the earlier epoch profiles are not well determined and the line
intensity values are approximate.  Although the signal-to-noise is
lower, clearly the form and evolution of the strong He\,{\sc i} 1.083~$\mu$m
profile was similar to that of H$\alpha$.

\begin{table}
\centering
\caption{Evolution of the H$\alpha$ line}
\begin{minipage}{\linewidth}
\renewcommand{\thefootnote}{\thempfootnote}
\renewcommand{\tabcolsep}{0.15cm}
\begin{tabular}{cccccccc} 
\hline 
Epoch  & \multicolumn{5}{c}{Vel. (km s$^{-1}$)}& Flux\footnote{Mean Flux given in $10^{-16}$ erg s$^{-1}$ cm$^{-2}$ \AA$^{-1}$.} & I \footnote{Intensity given in $10^{-13}$ erg  s$^{-1}$  cm$^{-2}$.}   \\ 
(d)   &   \multicolumn{2}{c}{Blue}  &  Mid- &    \multicolumn{2}{c}{Red} &   &   \\
        &  Edge &  Peak  &  Peak  &  Peak & Edge &   & \\ \hline
249     & --7060 & --4860  & $\sim$0 &  3400 & $\sim$7000 & 11.8  & 3.52 \\
305.3   & --6400 & --4570  &  --560  &  3450 & 6050 &  8.4  & 2.30   \\
346.4   & --6120 & --4340  &  --380  &  3770 & 5460 &  6.2  & 1.57 \\
418.3   & --6400 & --3930  &  --650  &  3860 & 6050 &  4.3  & 1.17  \\ 
474.3   & --6260 & --3890  &  --560  &  3950 & 5590 &  2.07 & 0.54  \\
658     & --5610 & --3940  &  --550  &  3540 & 5270 & (1.0)\footnote{Figures
in brackets were estimated by interpolation} & (0.25)  \\ 
1093.4  & --5760 & --3430  &   --60  &   --  & 5460 &  0.10 & 0.027 \\ \hline
\vspace{-0.8cm} 
\end{tabular} 
\end{minipage}
\end{table}

\begin{figure}
\vspace{10cm} \includegraphics{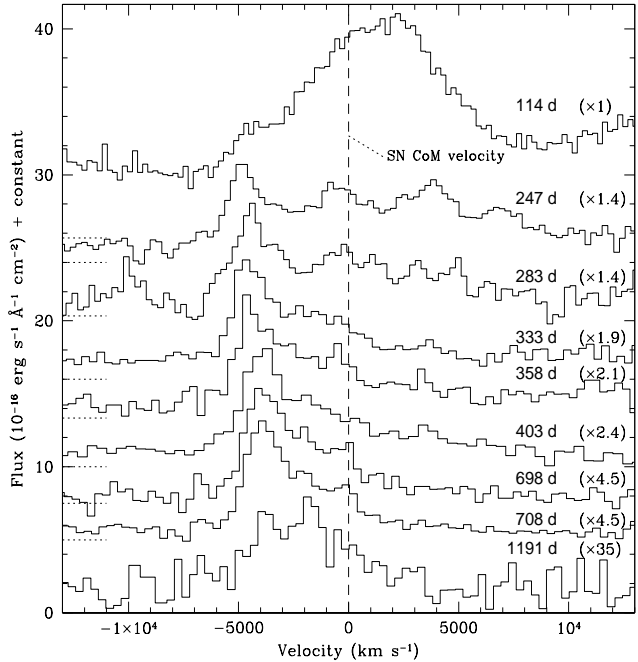}
\caption[]{Evolution of the He\,{\sc i}~1.083~$\mu$m profile, in velocity
space. Zero velocity corresponds to the rest frame of the SN~1998S
system (see text). The spectra have been scaled to show an
approximately constant height in the blue-shifted peak.  They have
also been displaced vertically for clarity.  The zero flux levels are
shown by the horizontal dotted lines on the left-hand side.  The 114~d
spectrum is from Fassia et al.  (2001) and the 247~d and 283~d spectra
are from Gerardy et al. (2000).}
\end{figure}

\section{Comparison of the IR spectral energy distribution with thermal continua}\label{bbfit}

The intensity and rise of the $HK$ continuum towards longer
wavelengths (Fig.~7) together with the relatively bright $L'$ and $M'$
magnitudes (Table~2) suggests strongly that the IR emission was due to
hot dust condensing in the ejecta and/or pre-existing in the CSM.  To
investigate the location and hence origin of the dust, we fitted the
IR $HKL'M'$ photometry flux values at each epoch (i.e. the IR SED),
with single-temperature blackbodies.  We excluded the $J$-band fluxes
from the fits since these were probably dominated by emission from the
ejecta gas (e.g. P$\beta$ emission, residual photospheric emission)
rather than hot dust (cf. Fig.~11).  It is possible that the $H$-band
might also have been slightly contaminated by non-dust emission.
However, approximate extrapolation of the $J$-band continuum (Fig.~11)
to longer wavelengths suggests that non-dust emission contributed less
than $\sim$20\% of the total flux in the $H$-band.  The magnitudes for
each waveband were dereddened using the empirical formula of Cardelli,
Clayton \& Mathis (1989) and adopting A$_V$=0.68 (Fassia et al. 2000).
Using the effective wavelengths of the filters, the dereddened IR
magnitudes were then converted into fluxes using the calibration curve
of Bersanelli, Bouchet \& Falomo (1991).  Using $\chi^2$ minimisation,
blackbody fits to the IR fluxes at each epoch were then performed,
with temperature T$_{bb}$ and solid angle as free parameters.  The
errors used in the fits were not just the statistical errors given in
Table~2, since these errors do not include systematic uncertainties
arising from varying atmospheric conditions from epoch to epoch or
between the target and standard. To obtain a more realistic assessment
of the true error we note that, up to $\sim$800~days, the $HKL'$ light
curves declined approximately linearly (magnitudes v. time) but with
an apparently random scatter in the measured values about this line.
We assume that this scatter is due to the combination of statistical
and systematic errors, rather than real fluctuations in the supernova
decline.  From the scatter we derive total uncertainties of at least
$\pm$0.05 magnitudes in $HK$, and $\pm$0.06 in $L'$. Where the
statistical errors are larger than this, then these were used.  In the
$M'$-band the relatively small number of points and the apparent
sudden fading around 700~days (cf. Fig.~2) means that it is not
possible to estimate the total error from the scatter.  However, given
that the statistical error alone is $\pm$0.1 to $\pm$0.15~mags, we
adopt a total uncertainty of $\pm$0.2 in $M'$.  The fits were repeated
using a blackbody scaled by $\lambda^{-1}$ and $\lambda^{-2}$
emissivity dependence such as might be the case for optically-thin
emission.\\

\begin{table}
\centering
\caption{Evolution of the He\,{\sc i}~1.083~$\mu$m line}
\begin{minipage}{\linewidth}
\renewcommand{\thefootnote}{\thempfootnote}
\renewcommand{\tabcolsep}{0.3cm}
\begin{tabular}{cccccc} 
\hline 
Epoch  & \multicolumn{3}{c}{Vel. (km s$^{-1}$)}&   Flux\footnote{Mean Flux given in  $10^{-17}$ erg s$^{-1}$  cm$^{-2}$ \AA$^{-1}$.} & I\footnote{Intensity given in  $10^{-14}$ erg  s$^{-1}$  cm$^{-2}$.} \\ 
(d)   &   \multicolumn{2}{c}{Blue}  & Red &  &  \\
       &  Edge &  Peak & Edge &  & \\ \hline
247\footnote{The earlier epochs are slightly contaminated by P$\gamma$
emission and so the red edge is undetermined and the intensities
are only approximate.  }

    & -6760 & -4790 &  --  & $\sim$22 & $\sim$10 \\
283    & -6120 & -4380 &  --  & $\sim$20 & $\sim$8 \\
333.8  & -6370 & -4520 & -- & $\sim$13 & $\sim$5 \\ 
358.9  & -5180 & -4440 & 4700 & 13 & 4.8 \\
403.9  & -5820 & -3720 & 4710 & 12 & 4.5 \\
698.0  & -5040 & -3860 & 4600 & 6.5 & 2.2 \\
708.9  & -5210 & -3830 & 4510 & 5.9 & 2.1 \\
1191.6 & $\sim$-5000 & -1730 & $\sim$3500 & 1.0 &0.25 \\ \hline
\end{tabular} 
\end{minipage}
\end{table}

\begin{figure*}
\vspace{13.5cm} \includegraphics{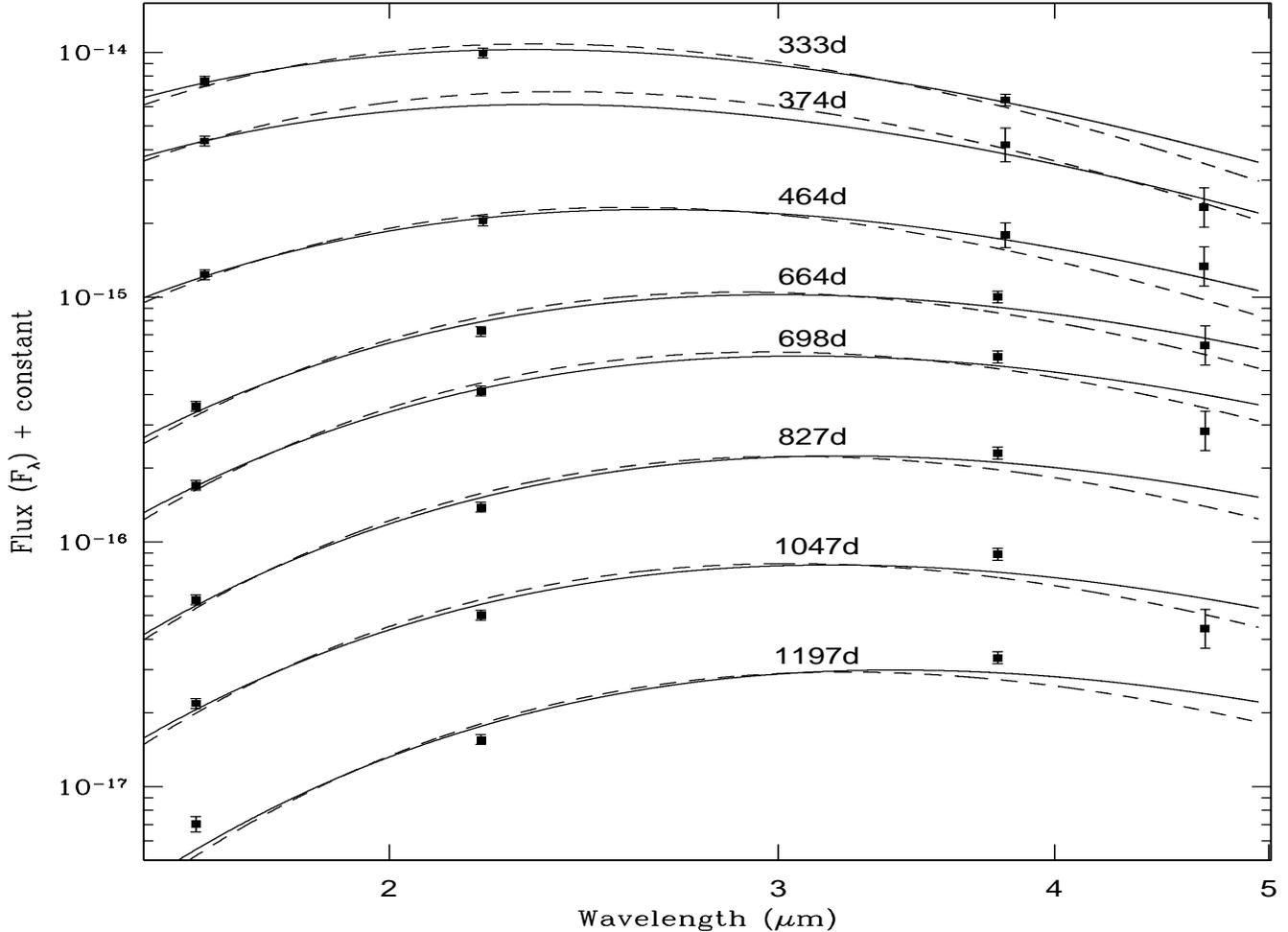}
\caption[] {Blackbody (solid curves) and emissivity $\lambda^{-1}$
fits (dashed curves) to $HKL'M'$ photometry of SN~1998S. The errors
shown include statistical and estimated systematic errors (see text
for details). The latest $L'$-band point was obtained by interpolation
using the $L'$ light curve. The observations are de-reddened, assuming
A$_V=0.68$ and the extinction law of Cardelli et al. (1989).}
\end{figure*}

\begin{table*}
 \centering
 \begin{minipage}{160mm}
  \caption{Details of fits to the NIR SED of SN~1998S} 
  \renewcommand{\tabcolsep}{0.2cm} 
  \begin{tabular}{@{}rlrrcccccc@{}}
 \hline
 Epoch &  npts(filters)\footnote{Number of points used in the blackbody fits (corresponding wavebands as given in brackets).} & $\chi^2_{red}$\footnote{Reduced chi square values.} &$\chi^2_{red}$\footnote{These illustrate the poorer fits achieved with a $\lambda^{-1}$-weighted blackbody law.} & T$_{bb}$\footnote{Fit parameters were derived after dereddening using Av=0.68 (Fassia et al. 2000) and the interstellar reddening law of Cardelli et al. (1989).} &  T$_{\lambda^{-1}}$  & L$_{bb}\footnote{The overall level of the luminosity light curve is
subject to an additional uncertainty of about $\pm$12\% due to distance uncertainty (see text).} 
$    &  r$_{bb}$       &  v$_{bb}$   & L ($^{56}$Ni)\footnote{Total power output of radioactive decay of 0.15~M$_{\odot}$ of $^{56}$Ni (Li, McCray \& Sunyaev 1994; Timmes et al. 1996; Fassia et al. 2000).} \\
(d)~~  &                & (bb)      & ($\lambda^{-1}$) & (K)  & (K) & ($10^{41}$\,erg\,s$^{-1}$) & ($10^{14}$\,cm) & (km\,s$^{-1}$)  &  ($10^{38}$\,erg\,s$^{-1}$)    \\\hline
 333.0 & 3($HKL'$)  & 0.7       &    5.6   & ~1254(23)\footnote{Internal errors in last one or two significant figures are shown in brackets.} & 1034(4) & 2.51(8)    & 119(6)   & 4150(210) & 995  \\
 374.0 & 3($HL'M'$) & 0.5       &    0.1   & 1234(16) & 1007(22)& ~2.42(11)   & 121(5)   & 3740(170) &  689  \\
 464.2 & 4($HKL'M'$)& 0.4       &    2.8   & 1102(22) & 941(1)  & ~2.01(12)   & 138(10)  & 3450(240) &  307  \\
 664.0 & 4($HKL'M'$)& 4.3       &    9.8   & ~970(12) & 836(5)  &  1.62(7)    & 161(7)   & 2800(130) &  52.3 \\
 698.2 & 4($HKL'M'$)& 3.6       &    4.5   & ~943(10) & 813(1)  &  1.49(6)    & 162(7)   & 2690(120) &  38.7 \\
 827.0 & 3($HKL'$)  & 7.7       &   18.5   & ~905(10) & 792(7)  &  1.21(6)    & 159(7)   & 2220(100) &  12.8 \\
1047.5 & 4($HKL'M'$)& 8.8       &   14.0   & ~915(10) & 795(7)  &  0.54(3)    & 104(5)   & 1150(50)~ &  2.35 \\
1197.8 & 3($HKL'$)& 23~         &   36.6   & ~856(11) & 749(7)  &  0.43(2)    & 106(6)   & 1020(60)~ &  0.98 \\ \hline 
\\ 
\vspace{-1.2cm}
\end{tabular}
\end{minipage}
\end{table*}

\begin{figure*}
\vspace{17.0cm} \includegraphics{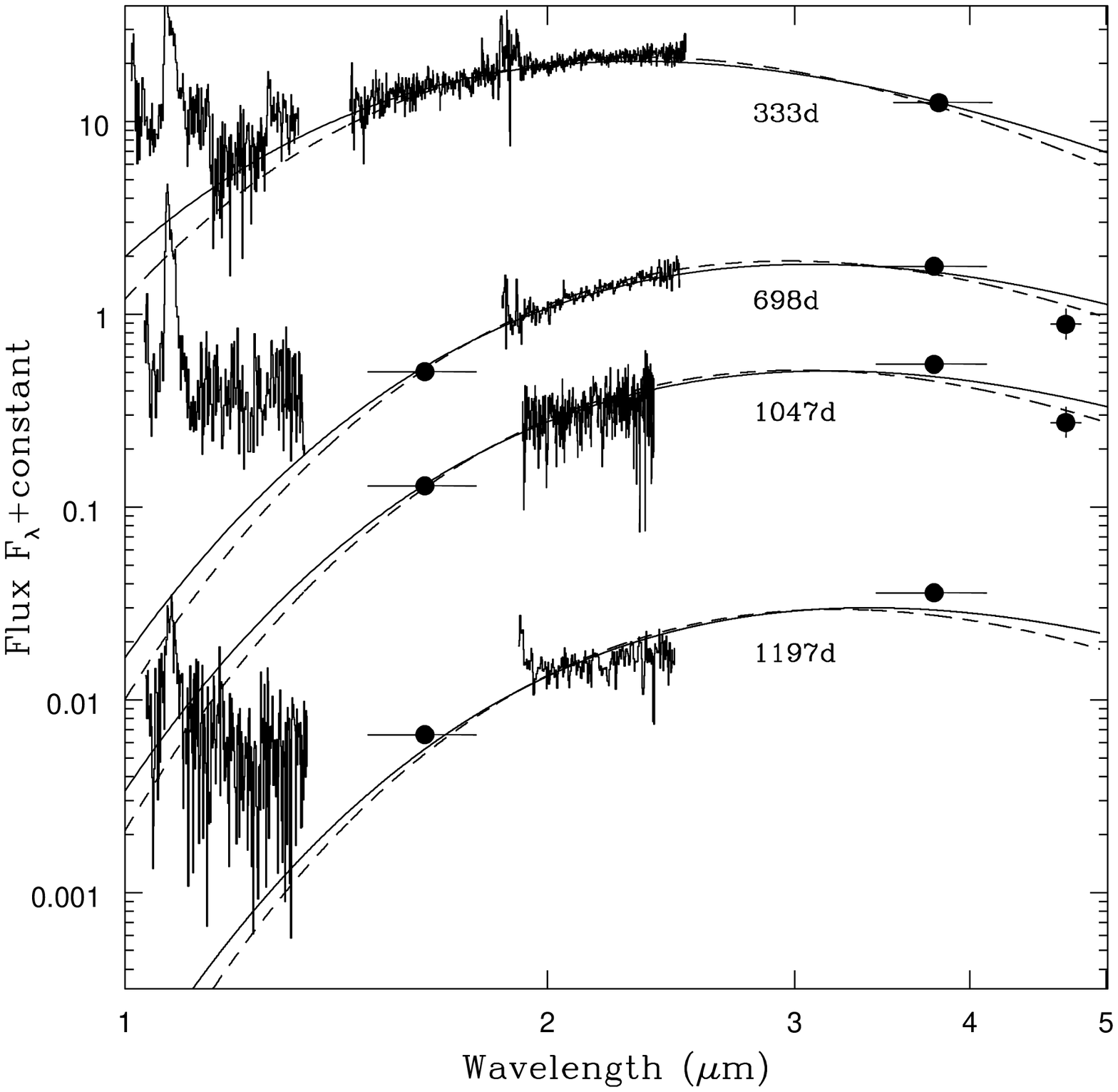}
\caption[]{Comparison of blackbody (solid curves) and emissivity
$\lambda^{-1}$ fits (dashed curves) given in Figure~10, with
contemporaneous IR spectra. Note that the fits did not include the
$J$-band photometry since this wavelength region was probably
dominated by emission from the ejecta gas rather than hot dust.  [The
lowest $J$-band spectrum corresponds to 1191~d (cf. Fig.~7) and so
should be compared with the 1197~d fit].  Approximate extrapolation of
the $J$-band continuum to longer wavelengths suggests that non-dust
emission contributed less than $\sim$20\% of the total flux in the
$H$-band.  Where no spectra are available, we show the $H$, $L'$ and
$M'$ photometry values used in the fits (cf. Fig.~10). The horizontal
bars on these points indicates the filter bandwidth (50\% levels). The
observations are de-reddened, assuming A$_V=0.68$ and the extinction
law of Cardelli et al. (1989).  }
\end{figure*}

The fits for a pure blackbody and for $\lambda^{-1}$ emissivity
dependence are illustrated in Fig.~10. Details of the fits are given
in Table~8.  Column~1 gives the epoch, and column~2 gives the number
of points used in the fit, together with the corresponding wavebands
in brackets.  The quality ($\chi^2_{red}$) of these fits is listed in
columns~3 (blackbody) and 4 ($\lambda^{-1}$).  For days 333, 374 and
464, single-temperature blackbody functions provide excellent
descriptions of the IR SED.  From day~664 the quality of the fits
becomes somewhat poorer. However, apart from 374~d, for all the epochs
the $\lambda^{-1}$ emissivity fits were somewhat less successful than
those using a pure blackbody SED.  The quality of fits with
$\lambda^{-2}$ dependence was even poorer.  In Fig.~11 we compare the
blackbody and $\lambda^{-1}$ emissivity curves with the coeval IR
spectra, when the wavelength coverage extended into the $K$-band.  It
can be seen that up to the end of year~2, the continua are
well-represented by the both pure blackbody and
$\lambda^{-1}$-weighted blackbody functions.  However, as Gerardy et
al. (2002) noted, $H$ and $K$ data alone are insufficient to
distinguish between different emissivity laws.  The pure blackbody
fits indicate that the temperature declined from $\sim$1250~K to
$\sim$950~K at the end of year~2, but then remained relatively
constant during the following year [cf. Table~8, col.~5 and
Fig.~12(b)].  This is reflected in the slowing of the reddening in
$K-L'$ (Fig.~3).  Not surprisingly, temperatures obtained from the
$\lambda^{-1}$ emissivity case are lower (Table~8, col.~6). The shape
and characteristic temperature of the SED supports our earlier
suspicion that the emission is produced by heated dust grains in
and/or around SN~1998S.  Similar conclusions were reached by Gerardy
et al. (2000).  However, our coverage to longer wavelengths allows us
also to conclude that pure blackbody fits appear somewhat superior.\\

Following a suggestion by the referee, we also examined fits using a
$\lambda^{+1}$ dependence, as might be the case if long iron whiskers
can form in the SN ejecta (see Wickramasinghe 1992; Hoyle \&
Wickramasinghe 1999).  We find that the quality of fits was generally
no better than for blackbodies and so on this basis alone we do not
rule out optically-thin emission with a $\lambda^{+1}$
dependence. However, a problem with such emission is that it would
require the condensing grains to be lying beyond the CDS.  This is
discussed further in Section 4.1.2.\\

\begin{figure}
\vspace{10.5cm} \includegraphics{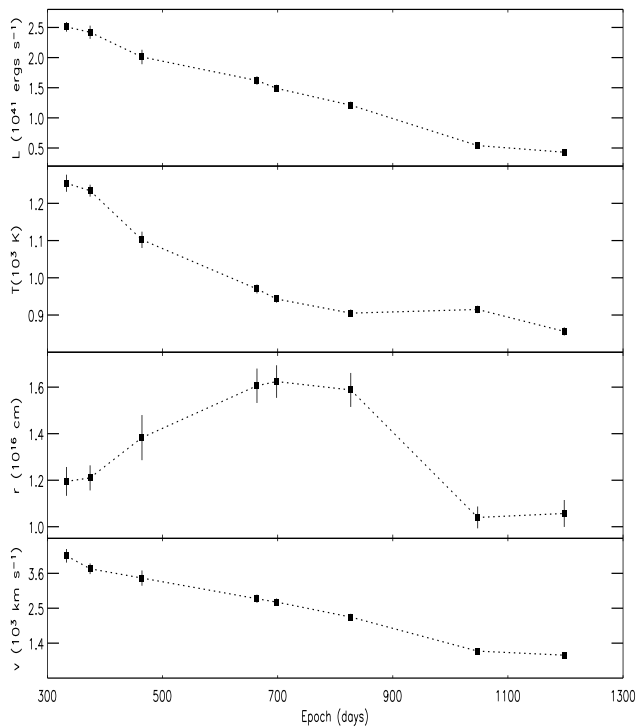}
\caption[]{Evolution of the luminosity, temperature, radius and
velocity derived from blackbody fits to the SN~1998S $HKL'M'$
magnitudes.}
\label{plot_bbquant_4panel}
\end{figure}

\begin{figure}
\vspace{9.5cm} \includegraphics{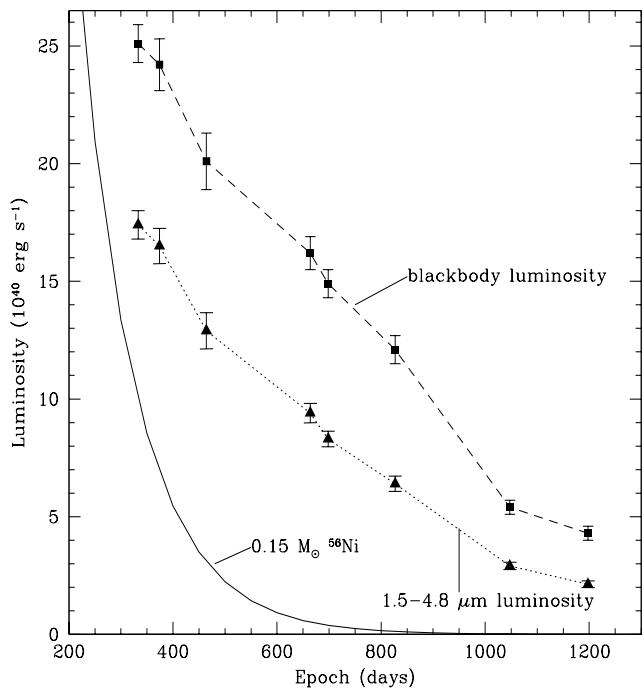}
\caption[] {Evolution of the total power output due to the radioactive
decay of 0.15~M$_{\odot}$ of $^{56}$Ni (Fassia et al. 2000) compared
with the IR luminosity obtained from (a) integrated blackbody fits
(dashed line) and (b) blackbody fits integrated over the range
1.5--4.8~$\mu$m (dotted line).}
\end{figure}

Adopting a distance of 17 Mpc (Tully 1988) and an explosion date of
1998 February 24.7 (see subsection 1.1), we found the blackbody radius
(r$_{bb}$) and velocity (v$_{bb}$) for each blackbody fit.  Finally,
we found the total flux and total luminosity (L$_{bb}$) by integrating
the best-fit blackbody curves. These values are listed in Table~8,
cols.~7--9.  and are plotted in Fig.~12(a,c,d).  Throughout the
333--1197d era, the blackbody fits indicate a monotonically declining
luminosity.  The derived blackbody velocity declined monotonically
from 4150~km s$^{-1}$ on day~333 to 1020~km s$^{-1}$ on day~1197,
while the radius peaked at $\sim$1100~AU at about 2 years.  Similar
conclusions would be reached for grey-body emission (e.g. clumps of
dust), except that the radii and velocities would be proportionally
higher.  For the $\lambda^{-1}$ and $\lambda^{-2}$ emissivity fits,
similar trends with time are obtained, but with much larger velocities
being required to account for the observed luminosity.  The
uncertainty in the extinction is quite large
(A$_V=0.68^{+0.34}_{-0.25}$, Fassia et al. 2000).  Therefore, to check
the effect on the above results, we repeated the fitting procedure
using the range of A$_V$ values encompassed by the errors.  We
conclude that the extinction uncertainty introduces an additional
error of only 1\% or 2\% in the derived parameters. 

As indicated in Section~1.1, there is some uncertainty about the
distance to NGC~3877 at the level of about $\pm$1~Mpc. Consequently
this introduces an additional uncertainty of about $\pm$12\%.  While
this does not affect the shape of the bolometric luminosity light
curve, it does act on its overall level.

\section{Discussion}
 
\subsection{Origin of the IR emission and location of the dust}

What powered the post-300~d IR emission from the dust, and where was
the dust located?  Integrating over the single temperature blackbody
fits, the total energy emitted by the dust region between days 300 and
1200 was about $10^{49}$~ergs, or $0.6 \times 10^{49}$~ergs in the
1.5-4.8~$\mu$m range.  Even for the more conservative value, this is a
factor of $\times50$ more than could be supplied by the decay of the
daughter products of 0.15~M$_{\odot}$ $^{56}$Ni (Fassia et al. 2001)
over the same period. This is illustrated in Fig.~13, where we compare
the radioactive luminosity with the contemporary IR luminosity derived
from the blackbody fits.\\

We can therefore immediately rule out radioactivity as the main source
of the post-300~d IR energy.  However, a much larger energy source is
available. It is likely that of order 10$^{51}$~ergs was stored in the
kinetic energy of the ejecta, so it would take only $\sim$1\% of this
to account for the IR emission through the radiative heating of
pre-existing (CSM) and/or newly-condensed (ejecta) dust.  For example,
for a progenitor with an extended structure or very dense wind, the
ejecta/CSM interaction could rather easily convert $\sim$10\% of the
kinetic energy to radiation (Falk \& Arnett 1977; Chugai et al. in
preparation), thus providing the energy of the early light curve
(0--40~days) which we know amounted to $\sim$10$^{50}$~ergs (Fassia et
al. 2000). If, then, a fraction (say 5--10\%) of this early light
curve emission were absorbed by CSM dust and re-emitted as an IR echo,
it would account for the observed IR radiation.  Gerardy et al.(2000)
suggested that dust could be heated as a result of energy released in
the ejecta/CSM interaction, either by direct shock-heating or by
absorption of X-rays from the interaction region.  However, they were
not able to address the origin of the dust, i.e. newly-formed in the
SN ejecta or pre-existing in the CSM.  As argued earlier, the large,
low-temperature $L'$-band flux seen at 136~days must have been due to
emission from pre-existing dust, i.e., an IR echo.  However, such an
argument is less convincing at the later times being considered here.
The blackbody fits produce velocities not exceeding 4200~km s$^{-1}$,
and generally slower than this (Table~8).  Grain condensation in
ejecta located at such velocities is conceivable.  However, while the
evolution of the line profiles provides convincing evidence that dust
did condense in the ejecta, this does not automatically mean that this
dust was also the source of the IR emission.  IR radiation from a
dusty CSM must therefore also be examined.

\subsubsection{IR emission from pre-existing CSM dust}
To try to distinguish between the IR echo and dust condensation
scenarios, we consider first the possibility that the IR emission
arose from pre-existing dust in the CSM.  It is unlikely that CSM dust
was directly heated by the supernova shock energy.  The supernova
radiation around maximum would have evaporated the dust to a radius
$r_v$.  The size of the dust-free cavity in such an explosion has been
estimated by a number of authors to lie in the range 4,000--80,000~AU
(Wright 1980; Bode \& Evans 1980; Dwek, 1983, 1985; Graham \& Meikle
1986; Gerardy et al. 2002).  Dwek (1983, 1985) finds that for a
supernova with a peak UV-optical bolometric luminosity of
$1\times10^{10}~L_\odot$ and an exponential decline rate timescale of
25 days, $r_v$ is $\sim$4,000~AU for carbon-rich grains
(T$_{evap}=1900~K$) and 20,000~AU for oxygen-rich grains
(T$_{evap}=1500~K$).  This is for 0.1~$\mu$m radius particles in an
$r^{-2}$ density distribution. The early bolometric light curve of
SN~1998S derived from blackbody fits to the optical-IR photometry
(Fassia et al.  2000) can be described as
$L=4.6\times10^{10}e^{-t/17.8d}~L_{\odot}$ erg s$^{-1}$.  The more
conservative spline-fit integration values yield
$L=1.5\times10^{10}e^{-t/24.4d}~L_{\odot}$ erg s$^{-1}$. Thus, given
that $r_v \propto (L_{peak})^{0.5}$ (Dwek 1985) we conclude that the
evaporation radius for SN~1998S would have been at least
5,000-9,000~AU for graphite grains, and at least 24,000-40,000~AU for
silicate grains.  The fastest moving ejecta material ($\sim$10,000~km
s$^{-1}$, Fassia et al. 2001) would have reached only 7,000~AU by the
latest epoch (1242~d), almost certainly placing it well within the
dust-free cavity throughout the observations presented here. The
ejecta would, therefore, have been unable to interact directly with
CSM dust.  \\

It is likely that an X-ray precursor is produced as the forward shock
moves through the dust-free low-density gas in the cavity.  These
X-rays would eventually reach the dust and hence may provide the
energy needed for the IR emission. A concern might be that the X-rays
could be severely attenuated by the gas in the cavity (assuming the
unshocked gas in the cavity was not already fully ionized by the
initial X-rays). Assuming a steady mass loss rate of
$\sim5\times10^{-4}$~M$_{\odot}$ yr$^{-1}$ (see Section 1.1) we find
that the gas column density between the shock front and the edge of
the dust free cavity (assumed to be $\sim$25,000~AU) is
$1.6\times10^{-22}$cm$^{-1}$ at 333~d.  This corresponds to an optical
depth $<0.2$ for X-rays of E$>3$~keV.  Given the hard X-ray spectrum
observed ($kT\approx10$~keV, Pooley et al. 2002) we do not expect
strong absorption of X-rays within the cavity.  We note that X-rays
would also be produced by the reverse shock.  However, if it is
radiative early on this might produce a CDS (Chevalier \& Fransson
1985), causing additional attenuation of this X-ray component.  The
difficulty with the X-ray precursor scenario is that the observed
X-ray luminosity is much less than that of the contemporary IR
luminosity. At 674~days, measurements with Chandra indicate that
$L_X=10^{40}$~erg s$^{-1}$ (Pooley et al. 2002), and as indicated
above, we expect that these hard X-rays would only be weakly absorbed
within the cavity.  However, this luminosity is about $\times15$ lower
than the blackbody-derived IR luminosity observed at this time.  Light
travel time effects mean that, in general, the observed IR emission
would have been produced by the X-ray precursor at an earlier,
possibly brighter phase. For example, by 664~d, a shock travelling at
$\sim$10,000~km s$^{-1}$ would have reached about 3800~AU.  The
largest delay would be for X-rays emitted directly away from us.  This
would amount to a delay of about 60~d for a 5,000~AU radius dust-free
cavity to about 460~d for a 40,000~AU cavity. However, even a year
later the IR luminosity has declined by only a factor of 3, i.e., it
is still $\times$5 more luminous than the X-ray flux one year before.
A perhaps rather desperate solution might be to suggest that more than
$\sim$90\% of the X-rays are being transformed into IR radiation,
i.e., most of the X-rays are absorbed by the dust, yet leaving enough
to be detected by Chandra.  (We note the actual percentage could be
somewhat less if the true X-ray temperature is higher than the
$\sim$10~keV derived from the Chandra observations so that the X-ray
flux was significant above Chandra's energy limit.)  In such a
scenario, we would expect the X-ray and IR light curves to be
reasonably correlated.  Between days 674 and 1044 the X-ray flux
declined by about $\times$2 while the IR luminosity fell by $\times$3
in the same period.  Given the complications of light travel times
across the CSM, this might be consistent with an X-ray
precursor-driven IR luminosity.  However, a particular difficulty with
the X-ray precursor model is that in general, for solid material,
X-rays are more penetrating than UV/optical radiation (Wilms, Allen \&
McCray 2000).  As indicated above, to account for the low X-ray/IR
luminosity ratio we need the dust optical depth to the X-rays to be at
least 2.5, in which case the optical depth to UV-optical photons would
be even higher.  Yet, for the IR-echo mechanism to produce the
observed IR luminosity we require an optical depth to
UV-optical-photons of about 0.2 (see below). In other words, in the
case of the X-ray precursor mechanism, if we invoke sufficient dust
absorption to account for the low X-ray/IR luminosity ratio then the
IR-echo mechanism predicts an IR luminosity which exceeds the observed
value by a large factor.  Even if we adopt the lower $1.5-4.8$ $\mu$m
luminosity values, a similar conclusion is reached.  Gerardy et
al. (2000) left open the possibility that the IR flux might arise from
either direct shock heating or X-ray-precursor heating of pre-existing
CSM dust.  However, from the above discussion we believe that both
possibilities are ruled out. Direct shock heating is impossible as the
ejecta would not yet have reached the edge of the dust-free cavity.
X-ray precursor heating is probably ruled out since it implies an
IR-echo of much greater intensity than the observations could
support.\\

We now consider thermal emission following the heating of CSM dust by
supernova radiation emitted around the time of maximum light, i.e., an
``IR echo''.  Owing to the light travel time across the CSM, the IR
emission seen at Earth at a given time originates from a zone bounded
by ellipsoidal surfaces, with the axis coincident with the
line-of-sight. The thickness of this zone is fixed by the
characteristic width of the UV-optical light curve around maximum.
While the UV-optical ellipsoid is still partially within the
dust-free cavity, we expect the IR light curve to be relatively flat
(Dwek 1983; Gerardy et al. 2002). However, once the whole ellipsoid
has left the cavity, the IR flux declines.  Even at our earliest epoch
of 333~d, we see no sign of a plateau in the observed IR light curves,
indicating that SN~1998S was already beyond the echo plateau phase.
At 333~days the vertex of the ellipsoid would be 30,000~AU from the
supernova. This indicates that, if the IR-echo scenario is valid, the
cavity radius must have been less than 30,000~AU.  \\

The flux, spectral energy distribution and evolution of a
supernova/CSM IR-echo has been examined by a number of authors (Wright
1980; Bode \& Evans 1980; Dwek, 1983, 1985; Graham \& Meikle 1986;
Gerardy et al. 2002). Of particular interest here is the study of
SN~1979C by Dwek (1983).  This supernova seems to have been similar to
SN 1998S (cf. Liu et al. 2000). Both supernovae were unusually
luminous at early times and both exhibited a strong IR excess at late
times.  Dwek (1983) argues that the late-time IR emission of SN~1979C
was due to an IR-echo produced in a massive circumstellar wind.  In
addition, the early-time bolometric light curve adopted by Dwek (1983)
for SN~1979C is exponential with a 23~day timescale - very similar to
the spline-fit light curve for SN~1998S (see above and Fassia et
al. 2000).  The extinction towards SN~1979C was somewhat less than for
SN~1998S, while their distances were almost identical.  After
correcting for the difference in extinction ($\Delta A_V\sim0.25$), we
find that the early-time bolometric light curve of SN~1998S
(spline-fit) is $\times1.3$ more luminous, while its post-300~d
blackbody-derived IR luminosity is $\times1.6$ less luminous than the
corresponding values for SN~1979C. Scaling Dwek's results to SN~1998S
indicates that about 20\% of its early-time UV-optical output must
have been absorbed by CSM dust.  Applying Dwek's $\lambda^{-1}$
emissivity model to a wind velocity of 40~km s$^{-1}$ (Fassia et
al. 2001), this yields a mass-loss rate of $1.1\times10^{-4}M_{\odot}$
yr$^{-1}$, comparable to the values indicated by optical, X-ray and
radio studies (Anupama et al. 2001; Fassia et al. 2001; Pooley et
al. 2002). Dwek's model parameters include an ISM gas-to-dust ratio of
160, a dust-free cavity radius of 20,000~AU, a steady mass-loss phase
of 1.7~years, a grain radius of 0.1~$\mu$m and a grain material
density of 3~g~cm$^{-3}$.  A longer mass-loss phase would result in
only a modest decrease in the derived mass-loss rate.  \\

The $K$ and $L$-band IR-echo light curves of SN~1979C have been
calculated by Bode \& Evans (1980).  At about 1~year, they find
decline rates of 0.44~mag/100~d in $K$ and 0.3~mag/100~d in $L$.  Dwek
finds an $L$-band decline rate of 0.4~mag/100~d for a putative
galactic supernova once the plateau phase is over.  Both authors
assume $r^{-2}$ CSM density laws.  For SN~1998S, we observe decline
rates of about 0.35~mag/100~d in $K$ and 0.16~mag/100~d in $L'$.
Similarly, for the total IR luminosity decline rate, while Dwek
predicts an e-folding time of 340~days between years~1 and 2, our
observations indicate e-folding times of about 470~days ($1-5~\mu$m
luminosity) or 700~days (blackbody luminosity).  However, these
discrepancies may be indicative of a flatter CSM density law than
$r^{-2}$, i.e. that the CSM wind was stronger in the past.  In
addition, these models assumed a spherically symmetric wind, whereas
the SN~1998S CSM probably had a more flattened geometry (Leonard et
al. 2000; Gerardy et al. 2000). \\

In addition to the modest decline rate disagreement, we also note that
Dwek's model provides an imperfect match to the SN~1979C $HKL$ SED.
Similar discrepancies with the Dwek model are seen for the SN~1998S
SEDs. Dwek argues that this may be due to the contamination of the $H$
(and $J$) band fluxes with residual photospheric emission.  However,
our post-300~d IR spectra of SN~1998S do not support this explanation
(cf. Fig.~11, day~333).  We recall that the Dwek models utilise dust
emissivity laws of $\lambda^{-1}$ and $\lambda^{-2}$, whereas our
analysis indicates that single temperature blackbodies yield a
superior fit to the $HKL'M'$ SEDs of SN~1998S. This suggests that the
dust grains are in an opaque clumped distribution, or are of larger
individual size than in the Dwek model.  Of course, the SED of the
IR-echo model is made up of fluxes from material having a range of
temperatures.  It is therefore not clear that even with `black-body'
grains (or clumps) we would be able to reproduce the single
temperature behaviour.\\
  
We conclude that the IR-echo scenario is not ruled out as providing
the mechanism responsible for the post-300~d IR flux from SN~1998S.
However, to make it work it has to be pushed to quite an extreme case
where 20\% of the early-time light is absorbed by the dust.  There is
little sign of such absorption in the form of reddening although, as
already indicated, the dust grains could be clumped or are
individually too large to give a reddening effect in the wavelength
range studied.  In addition, the simple spherically-symmetric IR-echo
models do not fully agree with the observed decline rates and SEDs.
Further progress will probably require more detailed IR-echo
modelling, including asymmetric and episodic cases. \\

\subsubsection{IR emission from newly-condensed ejecta dust}
We now consider newly-condensed dust as the origin of the IR
luminosity.  Gerardy et al. (2000) suggested that the detailed shape
and evolution of the H$\alpha$ and He\,{\sc i} profiles could be due
to the ejecta impacting on a disk-shaped CSM, together with the
condensation of dust within the ejecta. We agree that this is a
promising scenario.  As both Gerardy et al. (2000) and Leonard et
al. (2000) point out, the relatively sudden fading of the central and
red-shifted components of the H\,{\sc i}, He\,{\sc i} line profiles
immediately suggests dust condensation in the ejecta, causing
obscuration of the central and receding regions.  A similar effect was
observed in the ejecta line profiles of SN~1987A (Lucy et al. 1989;
Spyromilio, Meikle \& Allen 1990; Danziger et al. 1991) and, more
recently, in the type IIP SN~1999em (Elmhamdi et al. 2003).  The
presence of CO emission from SN~1998S as early as 115~days (Fassia et
al.  2001) lends credence to the dust condensation scenario.
Moreover, in spite of the much lower S/N of the He\,{\sc
i}~1.083~$\mu$m profile, plus its contamination by P$\gamma$ up to
about the end of year~1, we judge that the obscuration effect is
comparably strong at 0.66~$\mu$m and 1.08~$\mu$m, at least up to about
day~400.  This suggests that the dust became quickly optically thick
either due to forming in clumps, or (less likely) growing rapidly to a
large grain size. After day~400, the very low S/N of the He\,{\sc i}
red wing makes comparison impractical.  \\

\begin{figure}
\vspace{9cm} \includegraphics{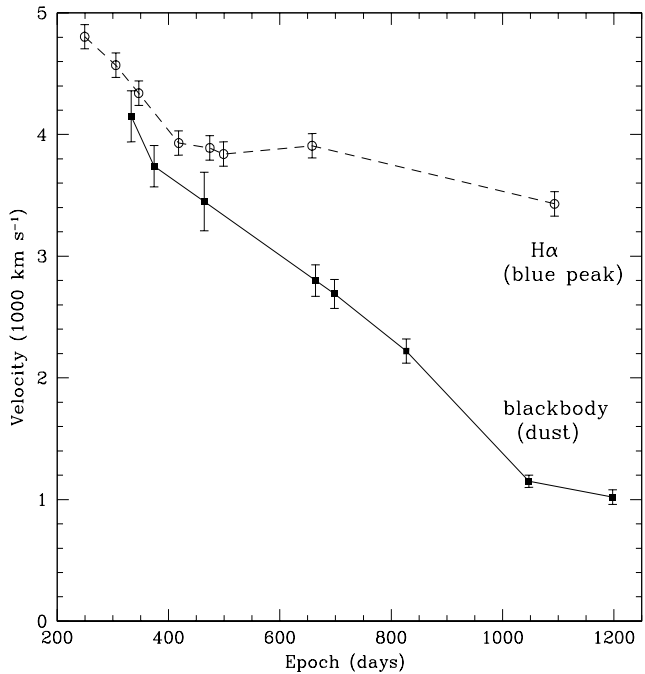}
\caption[] {Comparison of velocity evolution of H$\alpha$ blue peak
(Table~6), and of blackbody from fit to IR SED (Table~8).}
\end{figure}

Could newly-condensed dust also have been responsible for the
post-300~d IR luminosity of SN~1998S?  As already indicated, there is
an ample reservoir of energy available from the supernova kinetic
energy which could radiatively heat the dust.  For the pure blackbody
fits, we note that in the period $\sim$330--400~days the velocity of
the dust blackbody was $\sim$90\% of the velocity of the blue-shifted
H\,{\sc i}, He\,{\sc i} peaks (see Fig.~14).  Consequently, for
optically-thin dust characterised by $\lambda^{-1}$, $\lambda^{-2}$ or
$\lambda^{+1}$ emissivities to account for the observed IR luminosity
at this epoch, it would need to have condensed at velocities higher
than those indicated by the blue peaks of the H and He lines,
i.e. well beyond the main ejecta/CSM shock interaction.  This seems
unlikely.  Moreover, as pointed out already, $\lambda^{-1}$ and
$\lambda^{-2}$ emissivities yield poorer fits than does a pure
blackbody. We therefore discount IR luminosity characterised by
optically thin emissivities, and confine our attention to pure
blackbody emission.  After 333~d the dust and blue peak velocities
decline roughly at the same rate until about 400~days after which the
H\,{\sc i}, He\,{\sc i} peak velocity levelled off while that of the
dust continued to decline.  The near coincidence of the velocities for
the $\sim$100~days after 333~d suggests a possible physical link
between the origins of the H\,{\sc i}, He\,{\sc i} peaks and the IR
luminosity.  It has been recognised for many years (e.g., Chevalier \&
Fransson 1985) that the interaction of the supernova ejecta with a
dense CSM will produce outer and reverse shocks.  When radiative
cooling is important at the reverse shock front, the gas undergoes a
thermal instability, cooling to $\sim$10,000~K, thus forming a dense,
relatively cool zone - the CDS.  Line emission from low-ionisation
species in the CDS will be produced (Chevalier \& Fransson 1994).  We
suggest that this emission could have been responsible for the blue
and red peaks of the H$\alpha$ and He\,{\sc i} line profiles.  This
leads to the interesting possibility that dust may have formed in the
CDS at the ejecta/wind interface. (Note that this could be {\it as
well as} dust formation in the metal-rich SN ejecta interior.)  If
cooling in the outer layer of the CDS, shielded from the reverse shock
X-ray/UV radiation, brought the temperature to below the condensation
temperature, dust could have formed and survived there.  The physics
here is reminiscent of the radiative shock of colliding winds of
Wolf-Rayet stars, which is known to be a dust-forming site (Usov
1991). However, unlike the Wolf-Rayet material, the CDS has a normal
composition with the carbon-to-oxygen ratio being less than
unity. Consequently, at first sight, carbon dust might not be expected
to form since all the carbon in the CDS would be bound into CO.
However He$^{+}$ ions produced by X-ray absorption efficiently destroy
CO via the reaction He$^{+}+$CO $\rightarrow$ He+C$^{+}+$O (Lepp,
Dalgarno \& McCray 1990), yielding the relatively low observed mass of
CO (see Introduction).  We estimate that dust formed in the CDS after
day~200 may survive in the radiation field of SN 1998S.  The
occurrence of Rayleigh-Taylor or convective instabilities (Chevalier
\& Fransson 1994) can lead to the formation of opaque clumps of dust
within the CDS, producing significant occultation of the central and
receding parts of the supernova and radiative heating of the grains,
while at the same time allowing some of the line radiation to escape
from the approaching component of the CDS.  (Given that the CDS has a
temperature of $\sim$10,000~K, collisonal heating of the grains would
be negligible, e.g. Dwek \& Werner 1981).  Thus, this scenario might
simultaneously account for the strong IR flux, the obscuration effect
and the blackbody velocity coincidence with that of the line
profiles.\\

How much dust would be needed to achieve the required obscuration?
Consider dust grains of radius $a$ in spherical clumps of radius
$r_{cl}$.  For a spherical CDS whose thickness is small relative to
its radius, the total mass of dust in clumps, $M_{dust}$, required to
yield a CDS optical depth $\tau_{CDS}^{}$ is given by $M_{dust} =
(64/9)\pi^2 r_{CDS}^2 \tau_{CDS} n_g a^3 \rho
r_{cl}/(1-e^{-\tau_{cl}})$ where $r_{CDS}$ is the radius of the CDS,
$n_g$ is the grain number density and $\rho$ is the density of the
dust grain material. $\tau_{cl}$ is the effective optical depth of a
clump i.e. for a clump illuminated on one side, $e^{-\tau_{cl}}$ is 
the ratio of transmitted to incident fluxes, integrated
over the projected area of the clump, and is given by $e^{-\tau_{cl}} =$ 
$[1-(1+2\tau_0)e^{-2\tau_0}]/2{\tau_0}^2$, where $\tau_0$ is the centre 
to surface optical depth of a clump (e.g. Hobson \& Padman 1993).  
If we assume that the grains are sufficiently large that their
absorption cross-section at 5~$\mu$m equals their geometrical
cross-section, i.e. $a>\lambda/2\pi\approx1~\mu$m, then we may write
$\tau_0 = \pi a^2 r_{cl}\, n_g$. Substituting into the equation for 
$M_{dust}$, we obtain $M_{dust} =$ $(64/9)\pi r_{CDS}^2 \tau_{CDS} 
a \rho \tau_0/(1-e^{-\tau_{cl}})$. 
For smaller grains (cf. Kozasa, Hasegawa \& Nomoto 1989) if we assume 
that the ratio of absorption cross-section to geometrical cross section 
is proportional to $\lambda^{-1}$, then the dust mass becomes independent of 
$a$.  For larger grains, the mass would increase in proportion to the grain
radius.  At 346~d the radius of the H$\alpha$ blue-peak emission
region is $1.3\times10^{16}$~cm. At this epoch the dust blackbody
radius is 93\% of this. If we assume that the dust and H$\alpha$
emission zones are physically at about the same radius, then the
smaller dust blackbody radius can be interpreted as a covering factor
of about 86\%, or $\tau_{CDS}^{} \sim 2$. For near-opaque clumps, say
$\tau_{cl} > 2$ ($\tau_0>1.8$), $a=1~\mu$m, and a dust material
density of 2.5~g~cm$^{-3}$ (a typical density of grain material), we
obtain $M_{dust} > 2\times 10^{-3}$~M$_{\odot}$.  Note that this is a
lower limit i.e. larger values of $\tau_{cl}$ yield higher
masses. Repeating this calculation for subsequent epochs, we find that
the dust mass (lower limit) remains roughly constant up to about
700~d, after which it declines.  $\tau_{CDS}^{}$ declines
monotonically throughout this time, reaching 0.15 at 1000~d. \\

At 346~d the intensity ratio of the red and blue H$\alpha$ peaks is
0.25.  If we assume that the two peaks have the same intrinsic
luminosity at all times, then we can interpret the red deficit as
being due to an optical depth across the supernova of about 1.4. This
is similar to the optical depth of 2 derived in the previous paragraph
at the same epoch, using the velocities of the H$\alpha$ peak and the
IR blackbody. This lends support to the suggestion that condensing
dust is responsible for both the line profile attenuation {\it and}
the IR luminosity.  A difficulty is that while the optical depths
derived from the covering factor decrease with time, those derived
from the H$\alpha$ red/blue intensity ratio increase with time.  A
possible explanation might be that initially the optical depth was
dominated by hot, newly-formed dust and it was purely this dust which
was responsible for the obscuration {\it and} the observed near-IR
flux. This would explain the similarity at 346~d of the optical depths
derived in the two ways.  However, if dust condensation was continuous
throughout the 346--1093~d period it is possible that, if the dust
which formed at earlier epochs cooled, our NIR observations would have
detected a decreasing fraction of the total dust.  Such a process
would also mean that dust was present at an increasing range of
temperatures, and this may account for the poorer blackbody fits at
the later epochs.  In summary, we agree about the plausibility of the
dust-condensation scenario which Gerardy et al. (2000) put forward as
one of a number of possibilities.  We have refined this proposal in
that we suggest that the dust is formed in the CDS, with the energy
source for the IR emission coming ultimately from the shock
interaction. However we note that if, as discussed in Section 4.1.1,
the IR emission was predominantly due to an IR echo then the optical
depth `discrepancy' would disappear i.e. we would not then expect to
see a correlation between the optical depths derived via the covering
factor argument (which would be inappropriate for an IR echo) and
those derived from the H$\alpha$ red/blue intensity ratio.

\subsection{Line profiles at 3~years}

The H$\alpha$ and He\,{\sc i} 1.083~$\mu$m lines persisted beyond
3~years post-explosion (Figs. 8 \& 9). At 1093~days the extreme blue
wing of H$\alpha$ was at a velocity of $\sim$5800~km s$^{-1}$,
corresponding to 3650~AU, confirming that the CSM must have extended
to at least this distance (for a CSM velocity of 40 km s$^{-1}$,
this implies that the progenitor mass loss started at least 430 years 
prior to explosion). However, as explained above, even at this
late epoch it is unlikely that the ejecta would have encountered any
pre-existing CSM dust. The most notable change in the H$\alpha$
profile is the appearance of an emission feature peaking at about the
velocity of the supernova rest frame. Its wings extend to about
$\pm$1000~km s$^{-1}$ and so it cannot be due to the emergence of a
spatially-coincident H~II region as the supernova faded. It must be
physically connected to the supernova.  We note that the red side of
the H$\alpha$ feature is still very weak, presumably due to
attenuation by dust which condensed in the SN. Thus, it is unlikely
that we are witnessing emission from the central regions of the
supernova. An alternative explanation is indicated by the `clumpy
wind' model of Chugai \& Danziger (1994). In this scenario, we see
emission from individual slow-moving CSM clumps as they are engulfed
by the ejecta shock.  As mentioned above, this was suggested by
Gerardy et al. (2000) as an explanation for the earlier (and much
stronger) central peak of the H$\alpha$ feature. The increased
prominence of the central feature by day~1093 could be partly due to a
weakening of the direct interaction of the ejecta shock with the
undisturbed, inter-clump CSM which is responsible for the blue peak.
In the Introduction we pointed out that the early-time evolution of
the [O\,{\sc iii}]~5007~\AA\ CSM line indicated the existence of a
denser CSM beginning at a radius of 2850~AU. This distance would have
been reached by material moving at about 4500~km s$^{-1}$ by
1093~days.  We can see from Fig.~8 that at this epoch, a portion of
the blue wing of the H$\alpha$ line was moving at velocities exceeding
this value.  Thus, the H$\alpha$ luminosity at 1093~days may have been
enhanced by the interaction of the ejecta with a denser clumped CSM
region.\\

\begin{figure*}
\vspace{15cm} \includegraphics{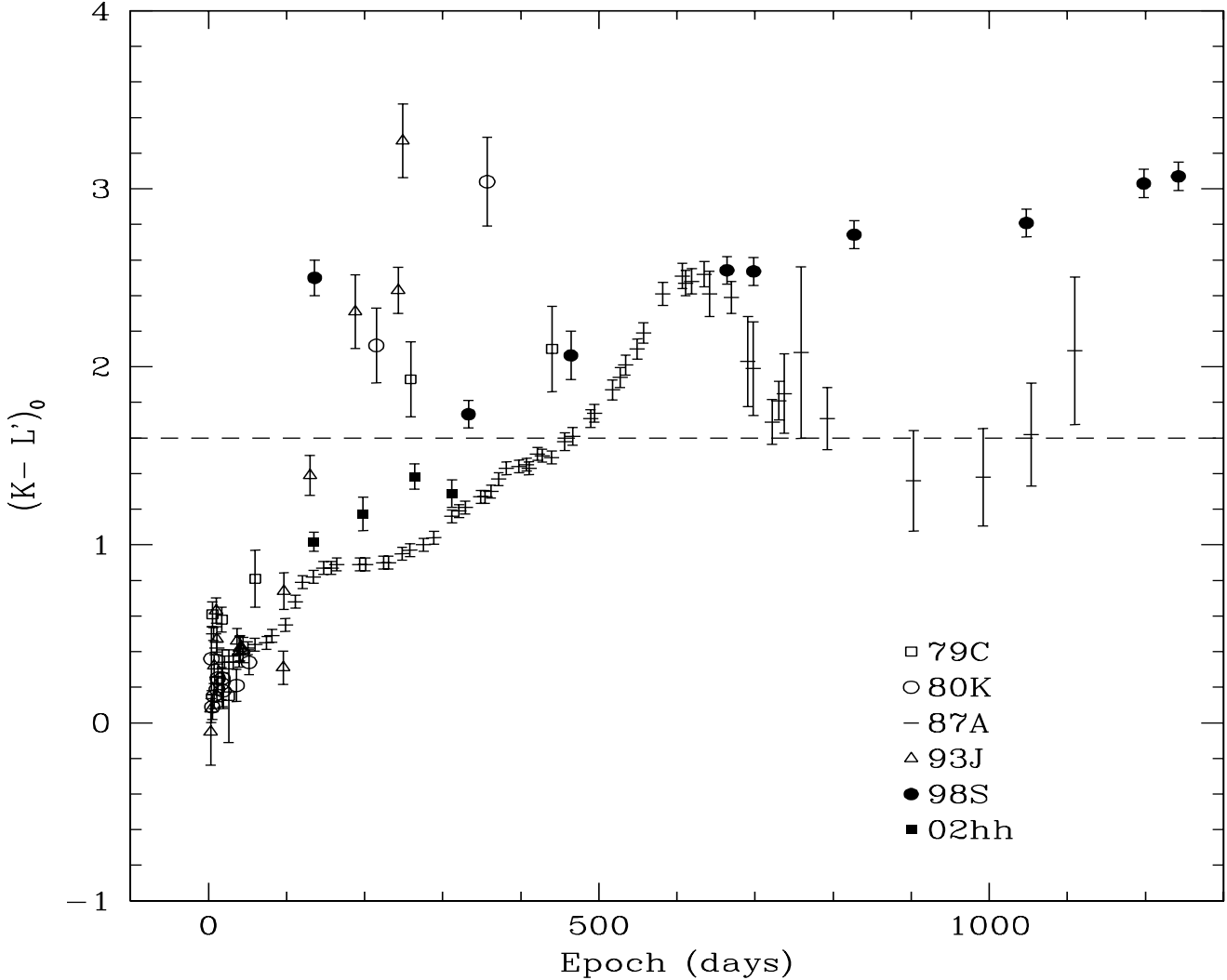}
\caption[]{Evolution of the extinction-corrected colour $(K-L')_0$ for
six core-collapse supernovae.  The epoch gives the approximate number
of days since outburst. We suggest that a value of $(K-L')_0>1.6$
(i.e. above the dashed horizontal line) indicates evidence of IR
emission from dust (see text).}
\label{plot_kl_CIT}
\end{figure*}

The He\,{\sc i} line is of much lower S/N and resolution.  It peaks
around --2000~km s$^{-1}$ with the wings extending to $\sim-5000$~km
s$^{-1}$ and $\sim+3500$~km s$^{-1}$. We find that by removing the
rest frame velocity component from the H$\alpha$ profile and then
degrading its resolution to that of the He\,{\sc i} line, a reasonable
match between the two profiles is obtained.  We conclude that, if
present, a rest frame zero velocity component was much weaker in the
He\,{\sc i}~1.083~$\mu$m line. \\

\subsection{$(K-L')_0$ colour and the origin of dust associated with supernovae}\label{nircolev} 

The work described above demonstrates the value of late-time
observations at wavelengths longer than the $K$-band for constraining
the conditions under which dust is present in or near supernovae.
Unfortunately, the only other SN for which such a comprehensive data
set is available is SN~1987A (Bouchet \& Danziger 1993 and references
therein; Caldwell et al. 1993 and references therein). Indeed, besides
SNe~1987A and 1998S, there are only four other SNe for which any $L$
or $L'$ observations beyond 100~days are available viz. SNe~1979C
(IIL) (Merrill 1980, private communication), 1980K (IIL) (Dwek et
al. 1983), 1993J (IIb) (Matthews et al. 2002) and 2002hh (IIp) (Pozzo
et al., in preparation).  In Fig.~15, we show the dereddened
colour, $(K-L')_0$, versus time for the 6 SNe.  For SNe~1979C and 1980K,
$L$-band rather than $L'$-band magnitudes were available.  In addition
the $L'$-band used for the SN~1993J observations is centred at a
somewhat shorter wavelength (3.68~$\mu$m) than the more typical value
of around 3.8~$\mu$m.  These magnitudes were therefore converted to
$L'$ (3.8~$\mu$m) as described in Appendix~A.  In addition, the
colours have been corrected for extinction using the Cardelli et
al. (1989) law (see Appendix~B).  For SNe~1998S and 1993J, the epochs
are with respect to outburst.  For the others, the fiducial time is BV
maximum.  However, the time from explosion to maximum is probably at
most 15--25 days, and so has a negligible effect on the conclusions
reached below since they are based on late-time observations.  For all
six SNe, we see (Fig.~15) that up to day~100, $(K-L')_0$ lies in the range
0.0--0.8, with a slight trend toward redder values with time.  Beyond
this epoch, the $(K-L')_0$ colour continues to redden, sometimes
dramatically so.  For blackbody emission, and a conservative maximum
dust temperature of 1300~K, $(K-L')_0=1.6$.  Moreover, we know that for
SNe~1987A and 1998S, when an IR continuum rising to longer wavelengths
was present, $(K-L')_0$ was also in excess of 1.6.  It therefore seems
reasonable to take $(K-L')_0>1.6$ as evidence of IR emission from dust,
via an IR-echo and/or from newly-condensing dust.  In Fig.~15,
$(K-L')_0$ values as high as $\sim$3 are seen, corresponding to a
blackbody temperature as low as 800~K. \\

During the 100--350~day era, while SNe~1987A and 2002hh showed a gradual
increase in $(K-L')_0$, the values for SNe~1979C, 1980K, 1993J and
1998S jumped to $(K-L')_0>\sim2.0$.  The different behaviour of
SNe~1987A and 2002hh on the one hand and the other 4 SNe on the other,
suggests IR emission was taking place under somewhat different
physical conditions in these two groups of SNe. A clue to the origin
of this difference is provided by observations which show that
SNe~1979C, 1980K, 1993J and 1998S were surrounded by large amounts of
circumstellar material (SN~1979C - Panagia et al. 1980; Fransson et
al.  1984; Fesen et al. 1999. SN~1980K - Leibundgut et al. 1991; Fesen
et al.  1999.  SN~1993J - Garnavich \& Ann 1994; 
Patat, Chugai \& Mazzali 1995; Matheson et al. 2000a, 2000b. SN~1998S -
Leonard et al. 2000; Gerardy et al. 2000; Fassia et al. 2001).  In
contrast, the wind density of the SN~1987A progenitor was very low in
its final blue supergiant phase ($<1\times10^{-8}$M$_{\odot}$yr$^{-1}$
for 500~km s$^{-1}$, Lundqvist 1999) and its red supergiant
wind with $<5\times10^{-5}$M$_{\odot}$yr$^{-1}$ (e.g., Blondin \&
Lundqvist 1993) is too far from the supernova to cause
early CSM interaction. Up to day~389 SN~2002hh also shows little
evidence of strong CSM interaction in its spectra (Pozzo et al., in
preparation).  We therefore suggest that supernovae which show
$(K-L')_0>1.6$ during 100--350~days post-explosion, probably have
massive circumstellar winds in their immediate vicinity.  \\

The precise origin of the strong IR excess [$(K-L')_0>1.6$] in the
100--350~d period is not straightforward to determine. As we have
already indicated, for SN~1998S at 136~d it is unlikely that the IR
emission is due to condensing dust since the required IR luminosity
and the red $(K-L')_0$ colour would require the emitting dust to be
lying at a velocity of $\sim$67,000~km s$^{-1}$.  A similar, though
somewhat weaker, point can be made for SN~1993J at 130~d where,
following correction for photospheric emission (Matthews et al. 2002)
the blackbody temperature would be $\sim$800~K and the blackbody
velocity around 8000~km s$^{-1}$. However, the situation is less clear
for the other examples of high $(K-L')_0$ in this era. The
exceptionally high late-time IR luminosity for SN~1979C poses a
particularly difficult challenge. For a blackbody to yield its $L'$
flux and colour on day~259 it would have to be expanding at 10,000~km
s$^{-1}$.  This seems to argue against an ejecta dust origin, although
we note the 223~d H$\alpha$ profile of SN~1979C also indicates a shock
velocity of about 10,000~km s$^{-1}$ (Filippenko 1997).  Yet to
account for the IR emission with the IR-echo model, it has to be
pushed to an extreme case where around 30\% of its early-time
UV-optical output must have been absorbed by CSM dust (Dwek 1983). It
may be, therefore that emission from both a CDS and an IR-echo have to
be invoked to account for the total IR emission from SN~1979C. For
SN~1998S there must also have been a period around 200--300~days when
the contributions from both sources were comparable. Given that
SNe~1980K and 1993J also possessed massive CSMs, we suggest that the
IR excess in these cases may also have been a mixture of emission from
CDS dust and CSM dust.  By $\sim$500~d we know that dust condensing
and cooling in the ejecta of SN~1987A was almost certainly responsible
for the reddening of the $(K-L')_0$ colour (e.g. Wooden 1989; Bouchet
\& Danziger 1993; Meikle et al. 1993; Roche et al. 1993; Wooden et al. 1993). 
From the discussion above we also favour
dust condensation in SN~1998S as the explanation for its IR luminosity
after $\sim$1~year.  Given the similarity of SN~1979C to 1998S, we
suggest that its day~440 $(K-L')_0$ point also indicates that most of
its IR emission was from CDS dust at this time. We note that at this
phase the blackbody velocity of SN~1979C would be only about 3000~km
s$^{-1}$. \\

We therefore propose that the appearance of a strong $(K-L')_0$ excess
during the first year is indicative of a massive CSM.  The IR flux can
be a mixture of emission from CDS dust and an IR-echo.  Determination
of the relative contributions requires good late-time IR and optical
coverage, as this paper has shown.  For the progenitors of SNe which
do not show an IR-excess in the 100--350~day period (e.g., SNe~1987A,
2002hh), it is much less likely that they were surrounded by a massive
CSM in their immediate vicinity.  If such SNe subsequently produce an
IR excess this is probably due to dust condensation throughout the
ejecta as was inferred for SN~1987A.

\section{Conclusions}

We have presented post-300~d infrared and optical observations of the
SN IIn 1998S which extend to longer wavelengths and later epochs than
has ever been achieved for this type of supernova. SN~1998S is only
the second ever supernova for which $M'$-band has been detected.  
The IR photometry presented here, together with an earlier
measurement by Fassia et al. (2000), has revealed a strong IR-excess
during the period 136 to 1242~days.  Spectroscopy in the $HK$ bands
show a continuum rising towards longer wavelengths.  Broad H$\alpha$
and He\,{\sc i} 1.083~$\mu$m line profiles present at $\sim$4~months
had undergone a dramatic change when they were recovered by Gerardy et
al. (2000) at about 8~months.  By this time, both lines took the form
of a triple peak structure with a central peak close to the supernova
rest frame velocity, and with the two other peaks lying at --4860~km
s$^{-1}$ and +3400~km s$^{-1}$ respectively. Gerardy et al. argue that
this could be indicative of emission from a disk-like or ring-like
CSM.  Our observations show that as the supernova evolved, the redward
half of the profiles became increasingly weak relative to the blue
part so that by the end of the second year the red side had almost
completely vanished.  We agree with Gerardy et al. (2000) and Leonard
et al. (2000) that this is strongly indicative of dust condensation
within the ejecta of the supernova.  The change in the appearance of
the H$\alpha$ profile by 1093~d together with the early-time evolution
of the narrow CSM [O\,{\sc iii}]~5007~\AA\ line may indicate a density
discontinuity at about 2850~AU implying that the progenitor mass-loss
rate underwent a decline about 340~years ago.  The velocity of the
extreme blue wing of the H$\alpha$ line at 1093~days implies that the
CSM extended to at least 3650~AU indicating that mass loss had been
taking place for at least 430~years prior to explosion. \\

The intensity and rise of the $HK$ continuum towards longer
wavelengths together with the relatively bright $L'$ and $M'$
magnitudes suggests strongly that the IR emission was due to hot dust
condensing in the ejecta and/or pre-existing in the CSM.  We find that
the SED represented by the $HKL'M'$ magnitudes and by the $HK$
continua (when available) can be well-reproduced with
single-temperature blackbody functions having temperatures of about
1250~K at 333~days declining to about 1100~K at 464~days.  Subsequent
epochs yield temperatures declining to about 900~K at 3~years,
although the fits are somewhat poorer.  This probably indicates an
increasing range of temperatures in the emission regions.  Fits
achieved with blackbodies weighted with a $\lambda^{-1}$ or
$\lambda^{-2}$ emissivity are almost always less successful than those
yielded by pure blackbodies. Newly-condensed iron whiskers having a
$\lambda^{+1}$ emissivity also seem unlikely as they would have to form
beyond the CDS.  We conclude that the IR emission arises from dust
in the form of IR-opaque clumps or having a large ($a>1\mu$m) grain
size. \\

We have considered possible origins for the strong, post-300~d IR
emission.  A significant amount of direct heating of condensed ejecta
dust by ongoing radioactive decay is ruled out since the observed IR
emission exceeds the radioactive luminosity by a substantial factor.
We also rule out the possibility of direct heating of pre-existing CSM
dust by the shock, since it would not have reached the edge of the
dust-free cavity.  X-ray precursor heating of CSM dust is also ruled
out since it would imply the additional presence of an IR-echo of much
greater luminosity than the observations could support.\\

We are therefore left with the IR-echo and condensing dust hypotheses.
In favour of the IR-echo explanation, Dwek's (1983) model with a
$\lambda^{-1}$ dust emissivity suggests that the observed IR
luminosity could be produced by a slow wind of
$\sim$10$^{-4}$M$_{\odot}$~yr$^{-1}$ which is similar to that
indicated by optical, X-ray and radio observations of SN~1998S.  This
assumes the typical ISM gas-to-dust ratio of 160 in the progenitor
CSM.  In addition, if we adopt a $\lambda^{-1}$ emissivity law, then
condensing dust would be unable to supply sufficient IR luminosity.
However, our measurements tend to disfavour SEDs weighted with a
$\lambda^{-1}$ or $\lambda^{-2}$ emissivity.  Moreover, for the
IR-echo scenario to work for SN~1998S at $t>300$~d, we must invoke a
rather extreme case where at least 20\% of the UV-optical luminosity
is absorbed by the CSM dust.  In addition, against the IR-echo
scenario is the disagreement in the observed and predicted decline
rates and SEDs, although it may be possible to eliminate these
discrepancies through more detailed IR-echo modelling, including
asymmetric and episodic cases.  We therefore do not rule out the
IR-echo scenario.  At the earlier epoch of 136~d, an IR-echo seems to
be the only plausible explanation for the IR luminosity. \\

In favour of the condensing dust hypothesis is the success of the
single temperature pure blackbody fits at about 1~year. Dust
condensing in optically thick clumps within a physically relatively
thin shell symmetrical about the supernova centre-of mass might be
expected to exhibit such temperature uniformity.  (Light travel time
effects would be negligible in this case since even at 2~years the
ejecta radius would only be about 10~light days.)  Moreover, the
coincidence at about 1~year of the blue-peak velocities of the
H$\alpha$ and He\,{\sc i} 1.083~$\mu$m lines with that derived from
the blackbody fits to the IR magnitudes suggests a physical
connection.  This is supported by the similar values at about 1~year
for the optical depths derived from the H$\alpha$ line profile, the
He\,{\sc i}~1.083~$\mu$m line profile, and that obtained from the
blackbody-derived dust covering fraction.  We suggest that this
connection is the CDS formed as a result of the ejecta-CSM
interaction. It is possible that conditions in the CDS would allow the
condensation of dust.  Moreover the highest blackbody temperature we
derive is 1250~K, which is also believed to be compatible with grain
condensation, though only slightly higher temperatures may not be
(Gehrz 1988).  Assuming that all the dust at about 1~year is at the
same temperature, we estimate that as early as 1~year, over
$10^{-3}$~M$_{\odot}$ had formed.  The apparent decrease in the dust
mass after about 700~d together with the reduced success of the
blackbody fits to the IR SED may indicate that dust condensation was
actually still ongoing but that an increasing proportion had cooled to
temperatures where the bulk of the flux was emitted beyond NIR
wavelengths.  Given these arguments, we tend to favour dust
condensation as the origin of {\it both} the post-300~d IR luminosity
and the line profile attenuation.\\

It should be stressed that in the dust condensation scenario, the
total dust mass produced could be very much larger than the
$\sim\times10^{-3}$~M$_{\odot}$ lower limit. Most of the dust could
have been invisible due to the clump opacities being much higher than
that necessary to produce a blackbody spectrum, and increasingly large
amounts could have gone undetected due to the cooling of an
increasingly large proportion of the dust beyond the reach of the NIR
observations.  The possibility of very opaque clouds is also raised by
Elmhamdi et al. (2003), who obtain a somewhat weaker lower limit of
$\sim10^{-4}$~M$_{\odot}$ of dust for the type~IIP SN~1999em based on
the evolution of the [O~{\sc i}]~6300~\AA\ profile.  In addition, our
proposal that dust formed in the CDS does not exclude the possibility
that dust also condensed deep within in the metal-rich SN ejecta
interior.  Indeed the presence of a CO emission zone on day~115 at a
velocity of only 2200~km~s$^{-1}$ (Fassia et al. 2001) suggests that a
cool dust-forming environment deep in the ejecta might have developed.
Our dust mass lower limit for SN~1998S is consistent with the
$\sim$0.1--0.3~M$_{\odot}$ of condensed dust in SN~1987A deduced from
metal depletion (Lucy et al. 1991; Dwek et al. 1992; Spyromilio \& Graham 1992).
However, SN~1987A was peculiar.  Moreover, SNe~IIn make up only
10--15\% of all type~II events (Cappellaro \& Turatto
2001).  Of the three CCSNe (87A, 98S, 99em), only SN~1999em may be
described as `typical', but it is only one event.  It is therefore
premature to conclude that high rates of dust condensation occur in
all type~II events.  Consequently, the exact significance of SNe~II
for cosmic dust production is undecided.  For further progress to be
made, studies such as presented here should be carried out for a
statistically significant sample of supernovae, and extended to even
longer IR wavelengths. \\

Comparison of the $(K-L')_0$ evolution for a number of type~II SNe,
suggests that this may be a useful tool for discriminating between
supernovae whose progenitors possessed or lacked a massive CSM.  We
propose that the appearance of a strong $(K-L')_0$ excess
[$(K-L')_0>1.6$] during the first year is indicative of a massive CSM.
The IR flux at this time can be a mixture of emission from CDS dust
and an IR-echo i.e., two populations of dust may exist.  The
newly-formed ejecta dust would be responsible for the attenuation of
the red sides of the line profiles.  Determination of the relative
contributions by the two populations to the IR flux requires good
wavelength coverage at late-times, extending as far as the $L'$ and
$M'$ bands or even further to the mid-IR.  For the progenitors of SNe
which do not show an IR-excess in the 100--350~day period (e.g.,
SNe~1987A, 2002hh), it is much less likely that they were surrounded
by a massive CSM in their immediate vicinity.  If such SNe
subsequently produce an IR excess this is probably due to dust
condensation throughout the ejecta as was inferred for SN~1987A.

\section*{Acknowledgements}
We thank T. Dahl\'en, D. Farrah, D. Folha, P. Hirst, S. Leggett and
D. Lennon for carrying out some of the observations presented here. We
are also grateful to R. Fesen, A. Filippenko, C. Gerardy, D. Leonard
and T. Matheson for providing their spectra in digitised format.  We
also thank D. Leonard and T. Matheson for the use of their unpublished
spectrum of 1999 December 15.  We thank L. Lucy for helpful
discussions.  This work is based on observations collected at the
Isaac Newton Telescope (INT), La Palma, the Nordic Optical Telescope
(NOT), La Palma and the United Kingdom Infrared Telescope (UKIRT),
Hawaii.  The INT is operated on the island of La Palma by the Isaac
Newton Group (ING) in the Spanish Observatorio del Roque de los
Muchachos of the Instituto de Astrofisica de Canarias.  The NOT is
operated on the island of La Palma jointly by Denmark, Finland,
Iceland, Norway, and Sweden, in the Spanish Observatorio del Roque de
los Muchachos of the Instituto de Astrofisica de Canarias.  UKIRT is
operated by the Joint Astronomy Centre on behalf of the U.K. Particle
Physics and Astronomy Research Council. Some of the data reported here
were obtained as part of the ING and UKIRT Service Programmes.  MP is
supported through PPARC grant PPA/G/S/2001/00512.  PL and NNC are
grateful for support from the Royal Swedish Academy of Sciences.

\appendix

\section[]{Conversion of $K-L$ to $K-L'$ colour}

For SNe~1979C and 1980K, $L$-band rather than $L'$-band magnitudes were
available.  In addition the $L'$-band used for the SN~1993J
observations is centred at a somewhat shorter wavelength (3.68~$\mu$m)
than the more typical value of around 3.8~$\mu$m.  We therefore
converted these magnitudes to $L'$(3.8~$\mu$m), based on the colours,
spectra and dust temperatures derived for SN~1998S around the same
period.  We adopted $K-L'$(3.8~$\mu$m) = $K-L$(3.5~$\mu$m) + 0.3 for
SNe~1979C and 1980K, and $K-L'$(3.8~$\mu$m) = $K-L'$(3.68~$\mu$m) +
0.12 for SN~1993J.  At epochs earlier than 100~days, when the spectra
are photosphere-dominated, the differences between $K-L$ and $K-L'$
are small and so no corrections were made.

\begin{table}
 \centering
 \begin{minipage}{80mm}
  \caption{E($K-L'$) values for the type~II SNe plotted in Fig.~15.}
  \renewcommand{\tabcolsep}{1.0cm} 
  \begin{tabular}{@{}ccc@{}}
 SN   & E($B-V$) & E($K-L'$)  \\ \hline
1979C & 0.18\footnote{de Vaucouleurs et al. 1981.}  & 0.04 \\
1980K & 0.4\,\footnote{Burstein \& Heiles 1982.}      & 0.08 \\
1987A & 0.2\,\footnote{Romaniello et al. 2002.}       & 0.04 \\
1993J & 0.24\footnote{Clocchiatti et al. 1995.}     & 0.05 \\
1998S & 0.22\,\footnote{Fassia et al. 2000.}          & 0.05 \\ 
2002hh& $\sim$2.0\,\footnote{Meikle et al. 2002}      & $\sim$0.48\,\, \\ \hline
\vspace{-0.7cm}
\end{tabular}
\end{minipage}
\end{table}

\section[]{Extinction correction for the $K-L'$ colour.}

Prior to plotting the $(K-L')_0$ colour evolution in Fig.~15, correction
for extinction was applied using E($K-L'$) $=0.211$ E($B-V$),
derived from the interstellar extinction law of Cardelli et al.
(1989).  These corrections are listed in Table~B1. Apart from the case
of SN~2002hh, the corrections are small.

\bsp

\label{lastpage}

\end{document}